\documentclass[referee]{aa-5-mpa}
\usepackage{graphicx}
\usepackage{natbib}
\usepackage{amssymb}
\usepackage{txfonts}
\usepackage{natbib}

\newcommand{\Iso}[2]{^{#1}{\rm #2}}
\newcommand{\msun}{M_\odot}

\newcommand{\teff}{T_\mathrm{eff}}

\begin{document}

\title{New Asymptotic Giant Branch models for a range of
  metallicities}
\titlerunning{AGB model grid}

\author{A.~Weiss\inst{1} and J.W.~Ferguson\inst{2,1}}

\institute{Max-Planck-Institut f\"ur Astrophysik,
           Karl-Schwarzschild-Str.~1, 85748 Garching,
           Federal Republic of Germany
           \and
           Physics Department, Wichita State University, Wichita, KS
           67260-0032, USA
           }

\offprints{A.~Weiss; (e-mail: weiss@mpa-garching.mpg.de)}
\mail{A.~Weiss}

\date{accepted for publication in Astron.\ Astrophys.\ 14/07/2009}

\abstract{We present a new grid of stellar model calculations for stars on
  the Asymptotic Giant Branch  between 1.0 and $6.0\,\msun$. Our grid
  consists of 10 chemical mixtures with 5 metallicities between
  $Z=0.0005$ and $Z=0.04$, and with both solar-like and
  $\alpha$-element enhanced metal ratios for each metallicity. We treat  
  consistently the carbon-enhancement of the stellar envelopes by using
  opacity tables with varying C/O-ratio and by employing theoretical mass
  loss rates for carbon stars. The low temperature opacities have been
  calculated specifically for this project. For oxygen stars we use an
  empirical mass loss formalism. The third dredge-up is naturally obtained
  by including convective overshooting. Our models reach effective
  temperatures in agreement with earlier synthetic models, which included
  approximative carbon-enriched molecular opacities and show good
  agreement with empirically determined carbon-star lifetimes. A fraction
  of the models could be followed into the post-AGB phase, for which we
  provide models in a mass range supplementing previous post-AGB
  calculations. Our grid constitutes the most extensive set of AGB-models,
  calculated with the latest physical input data and treating
  carbon-enhancement due to the third dredge-up most consistently.}

\maketitle
\clearpage

\section{Introduction}
\label{s:intro}
A number of stellar model libraries have been developed to serve as
databases for applications in various fields of astrophysics. These
range from fitting evolutionary tracks to individual stars, to
isochrone matching for stellar clusters, up to complete population
syntheses of galaxies. The latter purpose is probably the most
frequent one. Examples for such libraries are those at Padova
\citep[and references therein]{gbbc:2000}, BaSTI at Teramo
\citep{pcsc:2004}, and most recently the one at Dartmouth College
\citep{dcjkbf:2008}. These libraries are constantly updated by new
calculations and extended by covering more and more chemical
compositions.

In terms of the evolutionary phases covered, all of the just cited libraries 
provide the results of full stellar models of low and intermediate
mass ($1 \lesssim M/\msun \lesssim 8$) up to the onset of thermal
pulses (TP) on the Asymptotic Giant Branch (AGB). One reason is
the prohibitive effort to follow the TP-AGB phase with
complete models, not to speak of the notorious numerical problems
encountered during this phase of strongly changing timescales that can
be as short as hours during a helium shell flashing. Another reason
not to include full models in the libraries is the uncertainty of AGB
calculations, owing to the importance of rather ill-known physical
effects, such as overshooting, rotation, and mass loss (see
\citealt{herwig:2005} for a review of the current state of the art problems). 

However, for populations with an age of a few hundred million to about
2~billion years, the contribution of intermediate mass stars cannot be
ignored. Because of their high luminosity they contribute significantly 
to the integrated light, and due to their low surface temperatures
they dominate the spectra and colours in the near infrared. 
\citet{lcg:2000}, \citet{rbcc:2005} and others have demonstrated
this effect convincingly. The Padova and BaSTI stellar
model libraries have included the TP-AGB recently
\citep{magi:2007,cpcs:2007} by making use of {\em synthetic
  AGB-models} \citep[for a selection of historical and modern
  synthetic AGB-models see][]{ibtrur:78,renvol:81,gdj:93,mbc:96}. 
Synthetic AGB-models try to predict basic stellar parameters, such as $L(t)$,
$\teff(t)$ and stellar yields without resorting to calculations of full
computations, but use relationships obtained from full calculations. 
However, they are by no means merely reproducing the full
models. Rather, they are using basic properties, such as the mass of the
helium core or the luminosity of the helium shell as function of time as input
for calculations of mass loss, effective temperature, envelope composition and
even nucleosynthesis at the bottom of the convective envelope (the {\rm Hot
  Bottom Burning}; HBB). The idea is to take the complicated core and helium
shell evolution from full models and add the evolution of the hydrogen layers,
which is less difficult to compute by on-line calculations. With this approach
the synthetic models can treat effects like the third-dredge up or mass loss
as an additional, free-to-chose property, which is then usually calibrated by
comparison of the synthetic models with observed AGB-star samples. 

To some extent all synthetic models depend on results from full
stellar modelling, and therefore extensive computations of the evolution along
the AGB are needed.
The most widely used AGB- and post-AGB tracks are those by
\citet{vasswood:93,vasswood:94} and \citet{bloecker:95a,bloecker:95b}. 
The former calculations were done for initial masses between 0.89 and
$5\,\msun$, and 
for chemical compositions of $Z=0.016$ (``solar''), 0.008 (``LMC''), 0.004
(``SMC''), and 0.001 (``Pop.~II''). 
However, not all the mass values were calculated for each metallicity. 
The most
metal-poor set consists of only the 1.0 and $1.5\,\msun$ model. Bl\"ocker
calculated models for only a Pop.~I metallicity, $Z=0.021$ ($1\leq M/\msun < 7$). 
\citet{waggroen:98} have provided fitting
functions for synthetic populations based on the set of AGB calculations by
\citet{wag:96} for three metallicities ($Z=0.02$, 0.008, and 0.001) and masses
from 1 to $7\,\msun$. Most recently, a new comprehensive set has been added by 
\citet{karakas:2003} for $1\leq  M/\msun \leq 6$ and $Z=0.004$, 0.008, and
0.02, although the intention of this work has been to follow the
nucleosynthesis in 
AGB stars, and not to provide a grid of AGB models for general
usage. Obviously, the emphasis in all work quoted has been to provide models
adequate for solar-type stars and for the Large and Small Magellanic Clouds,
from which we have the largest amount of information about this evolutionary
phase. For use in population synthesis models of galaxies, this range of
  model compositions is certainly not sufficient.

Many of the physical ingredients for the models have been improved, in part
considerably, since the generation of all these grids of AGB and post-AGB
models. Most importantly, the available models were calculated  using
opacities and equation of state prior than those by the OPAL group
\citep{ir:96,rsi:96} or by the Opacity Project \citep{seaton:2007}, which rests
on the ``MHD'' equation of state \citep{mhd:88}. Only in 
\citet{karakas:2003} high-temperature OPAL opacities were used. Similarly, the
nuclear reactions and neutrino emission rates for most grids are from the '80s
or are even older in some cases. 

It is therefore timely to provide new grids for AGB stars, which
include both the latest physical ingredients, and for a larger variety of chemical compositions. 
Sect.~\ref{s:program} presents the general
structure of the 
stellar evolution program used. In Sect.~\ref{s:progagb} the details
of the code specific to 
this project, which constitute the improvement over previous grids for AGB
models will be discussed in detail. Apart from the aspect of updated
constitutional physics and more extensive chemical compositions
our emphasis lies on a consistent treatment of carbon
enrichment of the envelope due to the third dredge-up. This includes the
influence on the opacities, which has been shown by
\citet{marigo:2002} to be crucial for the stars' temperature. This, in
turn, is the most 
important parameter for dust-driven mass loss, which dominates the late AGB
evolution and the transition to the post-AGB. In 
Sect.~\ref{s:calculations} the results of our calculations will be 
presented. Conclusions will close the paper in Sect.~\ref{s:discussion}.

\section{Program and calculation set-up}
\label{s:program}

\subsection{Chemical composition and stellar mass grid}
\label{s:grid}

In addition to the previously mentioned chemical compositions representative
for galactic Population~I (``solar''), the LMC and SMC stars, we have added one
super-solar composition with $Z=0.04$ and one metal-poor with $Z=0.0005$. For
the helium content of each mixture we used
\begin{equation} 
Y = Y_\mathrm{p} + Z \times \frac{\Delta Y}{\Delta Z} \, ,
\label{e:dydz}
\end{equation}
with $Y_\mathrm{p} = 0.245$ for the primordial helium content and $\Delta
Y/\Delta Z = 2.0$ for the connection between helium and metal production. The
first value is in agreement with the cosmological model and was also used, for
example, by \citet{cpcs:2007}, the second one is within the general range of
determination. In addition, for the first time $\alpha$-element enhancement
was taken into consideration in an AGB model grid, since 
the ratio of elements of the $\alpha$-group relative to those of the iron-peak
depends on the star formation history, which might differ radically from that
of the galactic disk and halo. Examples are the galactic bulge and elliptical galaxies,
where solar- and super-solar iron abundances along with $\alpha$-enhancement
are found. For each metallicity listed above, both a solar-scaled and an
$\alpha$-enhanced metal mixture was used. In Tables~\ref{t:1} and \ref{t:2} we
list ten mixtures for which we are providing models and the 
element distributions within the ``metals''. The solar mixture is that of
\citet{seaton:92}, which is almost identical to that of \citet{GN:93}; the
$\alpha$-enhancement is simply taken as an additional +0.4~dex for all
respective elements. These metal mixtures are identical to those in
\citet{coelho:2007}. 

\begin{table*}
\caption{Initial compositions used for the model calculations.}
\label{t:1}
\begin{tabular}{c|c|c|c|c|c|c|c|c|c|c}
\hline
metallicity & \multicolumn{2}{|c|}{super-solar}
& \multicolumn{2}{|c|}{solar}
& \multicolumn{2}{|c|}{LMC} & \multicolumn{2}{|c|}{SMC} 
& \multicolumn{2}{|c}{metal poor} \\
\hline
Z & \multicolumn{2}{|c|}{0.04} & \multicolumn{2}{|c|}{0.02}
& \multicolumn{2}{|c|}{0.008} & \multicolumn{2}{|c|}{0.004} 
& \multicolumn{2}{|c}{0.0005} \\
X & \multicolumn{2}{|c|}{0.635} & \multicolumn{2}{|c|}{0.695}
& \multicolumn{2}{|c|}{0.731} & \multicolumn{2}{|c|}{0.743} 
& \multicolumn{2}{|c}{0.7535} \\
Y & \multicolumn{2}{|c|}{0.325} & \multicolumn{2}{|c|}{0.285}
& \multicolumn{2}{|c|}{0.261} & \multicolumn{2}{|c|}{0.253} 
& \multicolumn{2}{|c}{0.2460} \\
\hline
met. scaling & sol. & $\alpha$-enh. & sol. & $\alpha$-enh.
& sol. & $\alpha$-enh. & sol. & $\alpha$-enh. 
& sol. & $\alpha$-enh. \\
\hline
mixture & I & II & III & IV & V & VI & VII & VIII & IX & X \\
\hline
\end{tabular}
\end{table*}

\begin{table}
\caption{Logarithmic element abundances within the metal group}
\label{t:2}
\begin{tabular}{r|c|c}
el.$^a$ &  solar$^b$  & $\alpha$-enhanced$^c$ \\
\hline
C & 8.55 & 8.55 \\
N & 7.97 & 7.97 \\
$^\ast$O & 8.87 & 9.27 \\
$^\ast$Ne & 8.07 & 8.47 \\
Na & 6.33 & 6.33 \\ 
$^\ast$Mg & 7.58 & 7.98 \\ 
Al & 6.47 & 6.47 \\ 
$^\ast$Si & 7.55 & 7.95 \\ 
P & 5.45 & 5.45 \\
$^\ast$S & 7.21 & 7.61 \\
Cl & 5.50 & 5.50 \\ 
Ar & 6.52 & 6.52 \\ 
K & 5.12 & 5.12 \\
$^\ast$Ca & 6.36 & 6.76 \\ 
$^\ast$Ti & 5.02 & 5.42 \\ 
Cr & 5.67 & 5.67 \\
Mn & 5.39 & 5.39 \\
Fe & 7.51 & 7.51 \\
Ni & 6.25 & 6.25 \\
\hline\noalign {\smallskip $^a$$\alpha$-elements are marked with an
  asterisk; $^b$\citet{seaton:92,GN:93}; $^c$an 
  $\alpha$-enhancement of +0.4~dex was assumed.}
\end{tabular}
\end{table}

For all 10 chemical compositions 11 initial mass values between 1 and
6~$\msun$ were calculated. They are 1.0, 1.2, 1.5, 1.6, 1.8, 2.0, 2.6, 3.0,
4.0, 5.0, and 6.0~$\msun$. Calculations were started on the zero-age
main-sequence (ZAMS) and continued as far as possible with the aim of
reaching the white dwarf cooling track.

\subsection{Stellar evolution program: basic properties}
\label{s:progbase}

For all calculations the Garching Stellar Evolution Code (GARSTEC)
as described most recently in \citet{wsch:2008} was used. Changes specifically
applied for the present calculations will be discussed in the following
section. The program is able to produce an up-to-date standard solar model  
\citep{wsch:2008}, from which a mixing length parameter of
$\alpha_\mathrm{MLT} = 1.74$ is obtained. For the calculations presented here
a value of 1.75 was used. The slight difference stems from the use of the
solar metal distribution of \citet{gs:98} in \citet{wsch:2008}.

GARSTEC is able to follow low-mass stars through the core helium flash
\citep[e.g.\ ][]{mw:2006} and many thermal pulses on the AGB without human
intervention since the numerical improvements by \citet{ww:94a}. 
Nevertheless, towards the end of the TP-AGB phase convergence problems
still exist and inhibit a continuous modeling of the whole stellar
evolution. We will return to this problem in Sect.~\ref{s:agbprop}.

Although particle diffusion is implemented in the program, and used in the
solar model calibration, the present calculations were done without employing
it. However, a diffusive scheme is used for convective mixing. Mixing and
nuclear burning are solved simultaneously in one set of equations, which is of
particular importance for fast nuclear burning phases, as in case of mixing of
protons into hot carbon-helium-layers. This may happen during the core helium
flash in extremely metal-poor stars \citep{scsw:2001} or in the TP-AGB phase.

Since we were interested in the structural properties of AGB stars, and not in
their chemical yields, the nuclear network was restricted to standard hydrogen
and helium burning reactions. If needed, both burning phases can be solved
together. The reaction rates are mainly from \citet{ADNDT:85} and
\citet{adel:98}, with the following updates: We use the  $3\alpha$-rate by
\citet{fynbo:2005}, the rate for the CNO bottleneck reaction
$\Iso{14}{N}(p,\gamma)\Iso{16}{O}$ by \citet{formicola:2004}, and the
$\Iso{12}{C}(\alpha,\gamma)\Iso{16}{O}$-rate by \citet{kunz:2002}. The
influence of the former two updates was discussed already in
\citet{wsksc:2005}. 
A deficit of this restricted nuclear reaction network is that
the hotter proton-cycles as well as $\alpha$-captures on intermediate 
elements such as nitrogen and oxygen are missing. Therefore the abundance of
nitrogen during the TP-AGB phase is always taken to be an upper limit, because
part of it could be processed into heavier elements by such reactions. 

The equation of state is the FreeEOS\footnote{available at {\tt
http://freeeos.sourceforge.net}} of A.~Irwin \citep[see][]{freeeos:2003}. We
  use Eddington grey atmospheres. All further details about our stellar
  evolution code can be found in \citet{wsch:2008}.

\section{Code improvements for AGB modeling}
\label{s:progagb}

The evolution along the AGB is characterized by the internal nuclear
processes, leading to increasing luminosities and larger stellar radii, and by
the strong mass loss due to stellar winds, which depend on mass, radius,
luminosity, and chemical composition of the envelope. As is known from the
initial-mass to final-mass relation \citep[see][for a comprehensive
  overview and Sect.~\ref{s:ifmr}]{weide:00}, the wide range of initial masses ($1 \lesssim M/\msun
\lesssim 8$) results -- due to the overwhelming effect of strong mass
loss at higher AGB-luminosities -- in quite a narrow range of white dwarf
masses ($0.5\lesssim M/\msun \lesssim 1.1$). The mass loss itself depends on
the chemical composition of the atmosphere and envelope, which itself is modified
by internal nuclear processes and mixing between the convective envelope and
regions of nuclear burning. Notably the third dredge-up, which leads to the
enrichment of the envelope with carbon from the helium-burning layers,
and which is the result of structure changes in the course of the thermal
pulses of the helium shell, is ultimately linked to the mass loss. The
enrichment of carbon in the outer stellar layers allows the formation of
carbon-molecules and dust, which then leads to dust-driven winds. Dust
formation and wind depend strongly on stellar temperatures, which itself
is determined by the superadiabatic convection of the envelope. 
The effectiveness of this convection depends upon the radiative
transport and opacites 
which in turn are a function of the carbon abundance.
It is the aim of our calculations
to treat the carbon abundance variations of the stellar envelope and the
consequences of it as consistently as possible. This implies a detailed
treatment of nuclear processes, mixing, opacities, and mass loss. In the
following we will present how we approach this problem. The nuclear
processes (carbon production during helium burning, CNO equilibrium)
were previously discused in the preceeding section.

\subsection{Dredge-up by convective overshooting}
\label{s:oversh}

In AGB-calculations with GARSTEC, as with most other programs, the third
dredge-up does not occur with the canonical physics described so far. If 
the third dredge-up occurs in our models it would show only for higher masses at high AGB-luminosities
(core masses) and low metallicity \citep{wag:96}. This is in contradiction to
observations and therefore additional mixing mechanisms are needed to
ensure the appearance of carbon stars on the lower AGB for all
populations. Several such processes have been invoked, such as gravity waves
\citep{dt:2003}, rotationally induced mixing \citep{lhws:99}, or convective
overshooting \citep{hbschee:97}. We adopt the latter approach here, but keep
in mind that overshooting could be just representative for the combined effect
of several physical processes acting simultaneously. In synthetic
models the amount of dredged-up material is adjusted to reproduce observed
carbon-star statistics.

The physical concept of overshooting we have implemented is that of an
exponentially decaying velocity field outside regions formally convective
according to the Schwarzschild-criterion \citep{fls:96}. This is cast into a
diffusion equation
\begin{equation} 
D_{\mathrm{ov}}=v_0 \cdot H_P(0)\exp\left[\frac{-2z}{f\cdot
    H_P(0)}\right] 
\label{e:ov}
\end{equation}
where $z$ is the distance from the boundary in the outer radiative
region, $H_P(0)$ is the pressure scale height taken at the boundary
  of the convectively unstable region, i.e.\ at $z=0$, and $v_0$ the
typical velocity of the convective elements (obtained from mixing length
theory) at the inner side of the Schwarzschild border, following
  \citet{hbschee:97}. 
$f$ is a
free overshooting parameter. It represents a measure of the efficiency
of the extra mixing. The larger the value of $f$, the further the extra mixing
extends outside the convective region.

The same approach has already been used, for example, by
\citet{herwig:2004a,herwig:2004b} for AGB models. The value for the
parameter $f$ initially used in \citet {hbschee:97} was $f=0.02$ -- based
on earlier main-sequence width fitting -- to obtain sufficient third
dredge-up. These authors stated that qualitatively their result do not
change if $f$ varies within a factor of 2. A somewhat smaller value of
$f=0.016$ was used by \citet{hbld:99} and repeatedly in later papers
\citep{herwig:2000,herwig:2004a,herwig:2004b}. 
However, detailed investigations into the nucleosynthetic products of
AGB-evolution with overshooting complicated the picture. While
\citet{hbld:99} emphasize the need for $f\approx 0.016$ for the
overshooting at the bottom of the pulse-driven convective layer, in
order to reproduce the abundance patterns of post-AGB stars of type
PG1159, \citet{lhlgs:2003} argue for a lower efficiency at the
convective border (they suggest a value of 0.008) to achieve better
agreement with detailed s-process abundance patterns. 
To further complicate issues, a recent
3-dimensional hydrodynamical study by \citet{herwig:2007} indicates a
varying effective $f$ at the bottom of the pulse-driven convection zone
between 0.01 and 0.14. 

Similarly, to
achieve sufficient efficiency of the $\Iso{13}{C}$ neutron source in
low-mass stars, \citet{lhlgs:2003} used $f=0.128$ for overshooting
from the bottom of the convective envelope, while \citet{herwig:2004b}
finds that a value of 0.03 prematurely stops the AGB evolution
of a $5\,\msun$ star. One should note that these detailed
investigations were done for either a single or a few stellar models, therefore
the results cannot be generalized to all masses and metallicities. 

In conclusion, given the unclear situation about the extent of overshooting
at the lower boundary of the pulse-driven convection zone, and
the finding that the overall evolution does not change dramatically if
$f$ is varied within a factor of 2, we chose to employ one value for
$f$ at all convective boundaries. This value is $f=0.016$, which is the
``generic'' value by Herwig.  For a slightly higher
value of $0.018$ we achieved good fits to galactic open clusters
colour-magnitude diagrams (unpublished). We are aware that our value
might be too high for the lower boundary of the pulse-driven
convection zone; a conclusion recently strengthened by 
\citet{sswm:2009} from implications of the initial-final-mass relation. 
We will return to this in Sect.~\ref{s:ifmr}.
Our choice is also in agreement with \citet{m3ba:2006}.

The only exception from our procedure is overshooting from
convective cores on the main sequence, where overshooting is restricted for
small convective core sizes, in agreement with similar approaches found in, for
example, \citet{vzmd:98}. 
For 1.0 $\leq M/\msun < 1.5 $ the overshooting efficiency is gradually
increased: starting from a value of 0 for $M_{\mathrm{ZAMS}} =
1.0 \msun$ it reaches 0.016 at $M_{\mathrm{ZAMS}} = 1.5 \msun$. 
The intermediate values are given by the
relation $f = 0.032\cdot(M/\msun - 1.0)$.

\begin{figure} 
\centerline{\includegraphics[angle=90,scale=0.45]{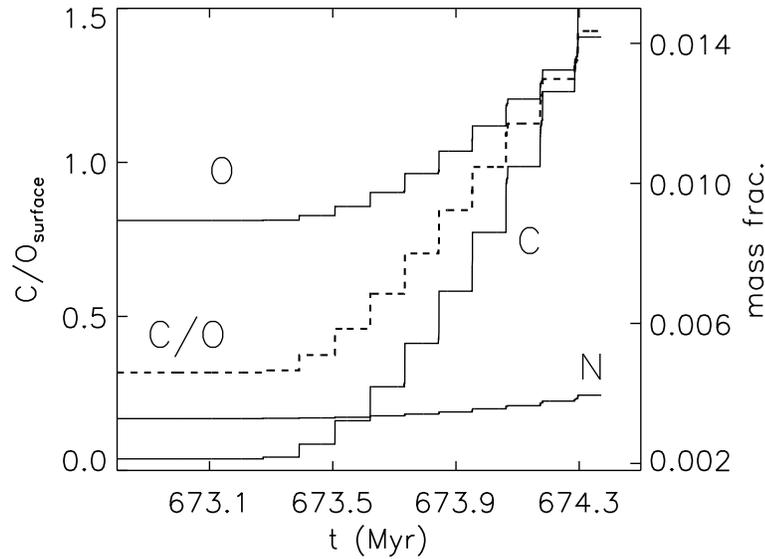}}
\caption{C, N, and O mass fractions (right axis) in an $2.6\,\msun$
  model ($Z=0.02$; solar metal 
  ratios) during   TP-AGB model as a consequence of the third
  dredge-up. The dashed line corresponds to the C/O-ratio (left axis).}
\label{f:3.1}
\end{figure}

Figure~\ref{f:3.1} shows the effect of the third dredge-up on the
abundances of C, N, and O and the C/O-ratio for a model of $2.6\msun$
with a solar-like metal abundance of 0.02 (mix III of Table~\ref{t:1}) as
caused by our overshooting description. This model does not 
experience HBB. Dredge-up starts after the 6th of 14 thermal pulses.
Note that oxygen is enhanced, too, due to the third dredge-up.

\subsection{Opacities for carbon-enriched compositions}
\label{s:opacities}

The carbon (and partially oxygen) enrichment of the envelope due to
the third dredge-up has to be reflected in the treatment of the
constitutional physics of the models. Where element abundances appear
explicitly, as in the nuclear reactions, this is trivial. The use of
tables for the equation of state and the Rosseland opacity inhibits
this direct approach, however. For the equation of state, composition
changes within 
the metal-group are not taken into account; generally, one assumes
that due to the low abundance of individual metals, even after
dredge-up, the equation of state is 
sufficiently accurate if the total metallicity $Z$ is taken into
account properly, which is the case in our calculations. 

The situation is different for opacities, where the absorption
properties can be more important than the absolute abundance of an
absorber. \citet{marigo:2002} has convincingly shown that the outer
envelope structure of AGB stars depends considerably on the opacities,
and that in particular carbon-enriched molecular opacities reduce
effective temperatures significantly, leading to much better agreement
of colours of synthetic populations with observations. 

Her adopted procedure to compute the
molecular opacities, through analytical fit relations, closely
resembles that of \cite{sc:75} and is incorporated in
the P.~Marigo synthetic code for TP-AGB evolution, and in the most
recent models of the Padova stellar model library \citep{magi:2007}. 
The possibility to
consistently compute the opacities for any chemical composition,
during the evolutionary calculations is a huge advantage of this
approach and the effects detected in the models help to account for a
number of observational properties of carbon stars. However, this has
never been implemented in full stellar models, with the exception of
the recent and independent work by \citet{csla:2007}, who used the
molecular opacities of \citet{la:2008}, but presented results solely
for one single model ($2\msun$; $Z=0.0001$). Recently, more $2\,\msun$
models for additional metallicities became available \citep{csgpdl:2009}.

\subsubsection{WSU tables for molecular opacities}
\label{s:wichita}

For our models new opacity tables have been prepared for C-enhanced
mixtures. For high temperatures, OPAL-tables for atomic opacities
\citep{ir:96} were obtained from the OPAL-website\footnote{\tt
  http://physci.llnl.gov/Research/OPAL}, and for low temperatures new
tables for molecular opacities were specifically generated with the
method and program described in \citet{fa:05}, in the following called
WSU (Wichita State University) tables. In all cases the
chemical compositions of low- and high-T tables agree and tables from
the different sources are combined as described in \citet{wsch:2008}.

\begin{table*}[ht]
\caption{Logarithmic metal abundances of the opacity tables for
  varying C/O ratios.} 
\label{t:3}
{
\begin{tabular}{c|c|c|c|c|c|c|c|c|c|c}
 & Set 2$^a$ & Set 3 & Set 4 & Set 5 & Set 6 & Set 8 & Set 9 & Set 10 & Set 11 & Set 12 \\
el. & C/O = 0.9 &
C/O = 1.0 & C/O = 1.1 & C/O = 3.0 & C/O = 20.0 & C/O = 0.9 & C/O = 1.0 & C/O = 1.1 & C/O = 3.0 & C/O = 20.0 \\
\hline
C  & 8.90 & 8.97 & 9.03 & 9.48 & 10.47 & 9.35 & 9.42 & 9.48 & 9.89 & 10.87 \\
N  & 8.14 & 8.18 & 8.22 & 8.56 & 8.91  & 8.45 & 8.50 & 8.56 & 8.92 & 9.29 \\
O  & 8.95 & 8.97 & 8.99 & 9.00 & 9.17  & 9.40 & 9.42 & 9.44 & 9.42 & 9.57 \\
Ne & 8.07 & 8.07 & 8.07 & 8.07 & 8.07  & 8.47 & 8.47 & 8.47 & 8.47 & 8.47 \\
Na & 6.33 & 6.33 & 6.33 & 6.33 & 6.33  & 6.33 & 6.33 & 6.33 & 6.33 & 6.33 \\
Mg$^b$ & 7.58 & 7.58 & 7.58 & 7.58 & 7.58  & 7.98 & 7.98 & 7.98 & 7.98 & 7.98 \\
\ldots & \ldots & \ldots & \ldots & \ldots & \ldots &\ldots & \ldots &
\ldots & \ldots & \ldots \\
\hline \noalign{\smallskip $^a$Sets 1 and 7 are omitted as they
  correspond to the standard solar-scaled and $\alpha$-enhanced
  distributions of Table~\ref{t:2}; $^b$table ends at Mg since
  elements heavier than oxygen remain as in the base mixtures.}
\end{tabular}
}
\end{table*}

For the absorption properties it is necessary to know which molecules are
present in cool stellar matter. A crucial quantity is the C/O-ratio,
because of the high binding energy of the CO-molecule. If the
$\mathrm{C/O}$ ratio is less than $\approx 1$ (oxygen-stars of type M or
S), stellar spectra show strong absorption bands of TiO, VO, and
$\mathrm{H}_2\mathrm{O}$. In the other case (carbon-stars of type C or R
and N), basically all O is bound in CO with CN, $\mathrm{C}_2$ and SiC and
some HCN and $\mathrm{C}_2\mathrm{H}_2$ formed from the remaining carbon.
Thus, the C/O-ratio is more
important than the absolute carbon abundance in the stellar mixture and
therefore the additional opacity tables were produced as function of
varying C/O-ratio (for each choice of $X$, $Z$, and $\alpha$-abundance),
but not of absolute carbon abundance. For the computations presented here
the table values of C/O were 0.48 (solar value), 0.9, 1.0, 1.1, 3.0, and
20. This choice was guided by an investigation of the change in the
Rosseland mean opacity due to variations of C/O.

\begin{figure}[ht]
\centerline{\includegraphics[scale=0.40,angle=90]{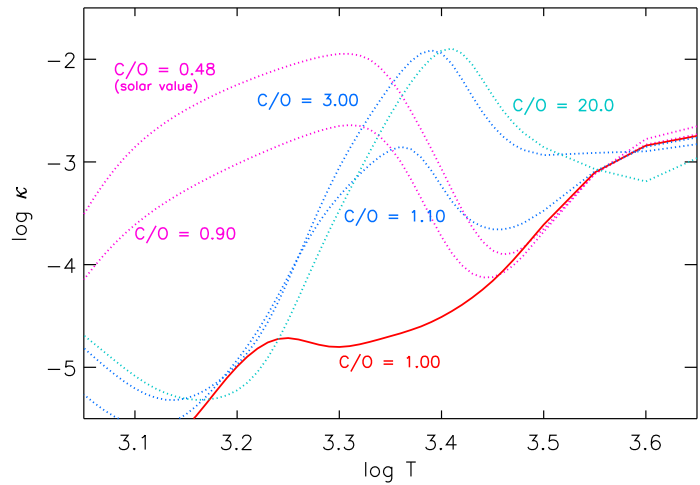}}
\caption{Rosseland mean opacity $\kappa$ as function of temperature for a
  mixture with $X=0.7$, $Z=0.02$ and a value of $\log R = \log \rho -
  3\log T+18 = -2.0$. The different lines are for the various C/O-ratios
  of the mixtures given in Table~\ref{t:3}.}
\label{f:3.2}
\end{figure}

Figure~\ref{f:3.2} shows how $\kappa$ changes with varying C/O
ratio.  The most sensitive regime is around $\mathrm{C/O} \approx 1$,
while for $\mathrm{C/O} >3$ hardly any change in the mean opacity is
apparent. The temperature range significant for AGB models is
approximately $\log T \geq 3.3$, which justifies our grid of C/O-ratios.
\citet{fd:2008} discuss in detail some of the important features of
Fig.~\ref{f:3.2}.  An important point is that as the C/O ratio increases
the amount of O available for molecular $\mathrm{H}_2\mathrm{O}$ is
decreased (becomes locked in CO) thus decreasing the mean opacity at
temperatures important for $\mathrm{H}_2\mathrm{O}$ absorption.  At 
$\mathrm{C/O} \approx 1$, most of the molecular opacity is in molecular
CO, a low absorber.  At higher values of C/O the opacity becomes dominated
by CN.

Recently, \citet{la:2009} published similar opacity data for chemical
mixtures with varying C and N abundances, which show similar behaviour as
our data (e.g.\ their figure 2 and discussion in \citet{fd:2008}).  In
particular, they stress that the regime around C/O=1 is the most sensitive
one that should be resolved well.

Table~\ref{t:3} lists the relative abundances of the metals for these
various C/O ratios, starting with the standard solar (Sets 2-6) or
$\alpha$-enhanced (Sets 8-12) composition, as given in
Table~\ref{t:2}. Note that not only carbon, but also oxygen and
nitrogen are enhanced. This is based on the fact that the third
dredge-up not only increases carbon, but also these elements to some
degree (see Fig.~\ref{f:3.1} for a typical case). The enhancements are
typical values. Heavier elements are 
assumed to remain unaltered, and therefore the table ends at
Mg.

To obtain the appropriate opacity for a point within a stellar model,
the following procedure was done: (1) select the set of tables for the
calculation, if the base mixture is either solar or $\alpha$-enhanced;
(2) interpolate within the $X$-dimension to the present hydrogen
abundance for all $Z$ and C/O-values; (3) interpolate to the present
$Z$ values; and finally (4) interpolate to the correct C/O-ratio.

\subsubsection{Approximative molecular opacities}
\label{s:marigo}

In addition to the low-temperature opacity tables discussed above, we
also had access to tables for carbon-enriched mixtures computed
 according to \citet{marigo:2002} and  kindly provided to us by
 P.~Marigo. Here, the 
carbon enhancement was simply added to the base solar composition,
beginning at $Z=0.02$. We used them for initial test calculations;
some of these will be compared below to our standard calculations in
which the tables of Sect.~\ref{s:wichita} were employed.

\begin{figure}[h]
 \centerline{\includegraphics[scale=0.6]{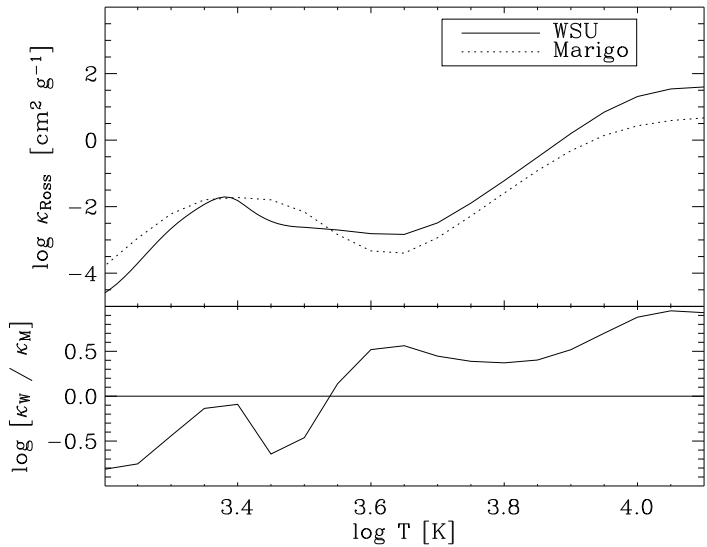}}
\caption{Comparison of the approximative molecular opacities by
  \citet{marigo:2002} with those by the WSU group (see
  Sect.~\ref{s:wichita}) for a  
  chemical composition with solar-scaled metallicity of $Z=0.04$,
  enriched by an additional amount of carbon of 0.06.}
\label{f:3.3}
\end{figure}

The availability of the approximative and the ``ab-initio'' tables
allows a brief comment on the quality of the approximation. We show
the comparison for a very metal-rich mixture of $Z=0.04$ ($X=0.7$)
with an additional amount of carbon of 0.06 (summing up to a total
metal fraction of 0.10) in Fig.~\ref{f:3.3}. The $\log R$-value is -3 in
this plot. The upper panel shows the Marigo and
WSU $\log \kappa$, the lower
panel shows the relative difference. The agreement is
indeed fairly satisfying, even for this extreme carbon
enhancement. The global behaviour as a function of temperature is
recovered by the approximative opacities, and in the interesting
temperature range $3.3 \lesssim \log 
T \lesssim 3.8$ the deviation remains below a factor 3 at low absolute
values. At temperatures above $\log T \gtrsim 3.8$ differences between the 
computations are not understood, however we use OPAL tables 
at these temperatures. Overall, we conclude that the Marigo
approximation is well suited to describe the basic effects of
carbon-enhancement on the temperatures of AGB stars. 

\subsection{Mass loss}

Mass loss is a decisive aspect of AGB evolution since it determines
how and when the TP-AGB phase ends, what yields can be expected from
intermediate-mass stars, and since it also influences possible nuclear
reactions at the bottom of the convective envelope. In the absence of
a complete theory for mass loss, simple mass loss formulas are
implemented in stellar codes. They are obtained by fitting either
empirical data or, if available, theoretical mass loss models. The
most widely known formulas used for AGB evolution calculations are
those by \citet{vasswood:93} and \citet{bloecker:95b}. In addition to
mass loss on the AGB, including the ``superwind'' phase, which
terminates this evolutionary phase, also the AGB--post-AGB transition
phase, the post-AGB evolution at increasingly higher $\teff$, and the
previous phases should be covered, although mass loss there is rather
insignificant when compared to that on the AGB.

In this investigation we employed the following mass loss
prescriptions:
As the basic formula, in particular for the RGB evolution, the
  Reimers relation \citep{reimers} is used. 

  \begin{equation}
    \dot{M}_\mathrm{R} = -4 \cdot
    10^{-13}\frac{(L/L_{\odot})(R/R_{\odot})}
    {(M/M_{\odot})}\eta_\mathrm{R}.
  \label{e:reim}
  \end{equation}

\noindent where $\eta_\mathrm{R}$ was set to 1.0 for stars
  with initial mass larger than $1.7\msun$, following
  \citet{bloecker:95b}. For models with 
  smaller ZAMS masses, a more standard value of 0.4 has been chosen as
  in \citet{karakas:2003}. In practice, the mass loss resulting from
  Eq.~\ref{e:reim} is insignificant with the exception of the RGB
  evolution of low-mass stars.

Once on the AGB, observed mass loss rates are higher than the standard
Reimers wind would indicate. The common picture is that the winds are
driven by radiation--dust interactions, where the dust production
itself is triggered or enhanced by radial pulsations
\citep[see][]{wallknapp:98,sedlwin:97}. At the present time, thorough
theoretical radiation-hydrodynamical models including dust production
are available only for carbon-rich chemical compositions, in which
nearly all oxygen is bound in CO, and the excessive carbon gives rise
to carbon-based molecules and dust. The Berlin group has published
both models and fitting formulas for such cases
\citep[e.g.][]{fgs:92,afs:97,wfbs:97}. 
We are employing here the mass loss rate by \citet{wswas:2002},
\begin{eqnarray} 
\log\dot{M}_\mathrm{AGB} &=& -4.52 +
    2.47\cdot\log\left(10^{-4}\frac{L}{L_{\odot}}\right) \nonumber \\
   &\phantom{=} & -6.81\cdot\log\left(\frac{\teff}{2 600 K}\right)
      -1.95\cdot\log\left(\frac{M}{M_{\odot}}\right). 
\label{e:wachter}
\end{eqnarray}
This formula does not include any dependence on the actual C/O-ratio
as did an earlier formulation by \citet{fleischer:94}, which was used
by \citet{wag:96}, the first work to include a mass loss rate based on
such models. \citet{wswas:2002} showed that the dependence of
the mass loss rate on the C/O-ratio is weak enough to be ignored in
comparison with all other uncertainties. This statement, however, was
obtained from investigating models at solar metallicity. In
\citet{wwss:2008} the same models were used to derive similar fit
formulas for LMC and SMC metallicities. Although the dependence on C/O
was again neglected, the coefficients in the equations corresponding
to Eq.~\ref{e:wachter} are different. We speculate that these coefficients also contain
the hidden effect that for a given C/O-ratio the absolute number of
unbound, therefore available carbon atoms differs between different
total metallicities. This is, according to \citet{mwhe:2008},
the decisive quantity. Since both papers appeared after our
computations were already performed, the metallicity dependence could
not be taken into account.

The second obvious
dependence on pulsation period has been implicitly included in the
$L$-term according to the period-luminosity relation by
\citet{gw:96}. 

For the case of oxygen stars, i.e.\ $\mathrm{C/O} < 1$, we used the
empirical fitting formula by \citet{vlczl:2005} obtained from
dust-enshrouded oxygen-rich AGB stars:
\begin{eqnarray} 
   \log\dot{M}_\mathrm{AGB} &=& -5.65 +
    1.05\cdot\log\left(10^{-4}\frac{L}{L_{\odot}}\right) \nonumber \\
    &\phantom{=}& -6.3\cdot\log\left(\frac{\teff}{3 500 K}\right).
\label{e:loon}
\end{eqnarray}
This rate is applicable only to stars with a pulsation period $P >
400$~days. Similar period cuts of 400 and 100~days have been employed by
\citet{vasswood:93} and \citet{bloecker:95b}. In accordance with the
latter reference, we use the \citet{oc:86} estimate of the pulsation
period for Mira variables of given mass and radius, and apply
Eqs.~\ref{e:wachter} and \ref{e:loon} if $P > 400$~days. For the carbon-rich
case, \citet{wswas:2002} actually give a lower critical luminosity,
but in our calculations it turned out that models reach the critical
pulsation period as oxygen-rich stars and only later become
carbon-stars at a higher luminosity. 

As a star leaves the AGB after a ``superwind'' has removed most of its
envelope, $\teff$ increases and the above formulae lead to a quickly
vanishing stellar wind. For this transition to the post-AGB phase no
suitable mass loss formula is available. \citet{detlef:07} argue,
based on hydro-simulations of dust envelopes around evolving post-AGB
stars, that the strong mass loss should pertain up to effective
temperatures of 5000 or 6000~K. We mimic this by keeping the AGB-wind
mass loss rates until $P$ has decreased to 150~days. From there to $P
= 100$~days, taken to be the beginning of the post-AGB phase, a linear
transition to the post-AGB wind is done. From there on, we again use
the Reimers relation (Eq.~\ref{e:reim}), or, if the rate is larger, the
radiation-driven wind formula already used by \citet{bloecker:95a},
and based on \citet{ppkmh:88}, namely:
\begin{equation}
    \dot{M}_{\mathrm{CSPN}} = -1.29 \cdot 10^{-15} L^{1.86}.
\label{e:cspn}
  \end{equation}

\begin{figure}[h]
\centerline{\includegraphics[scale=0.45,angle=90]{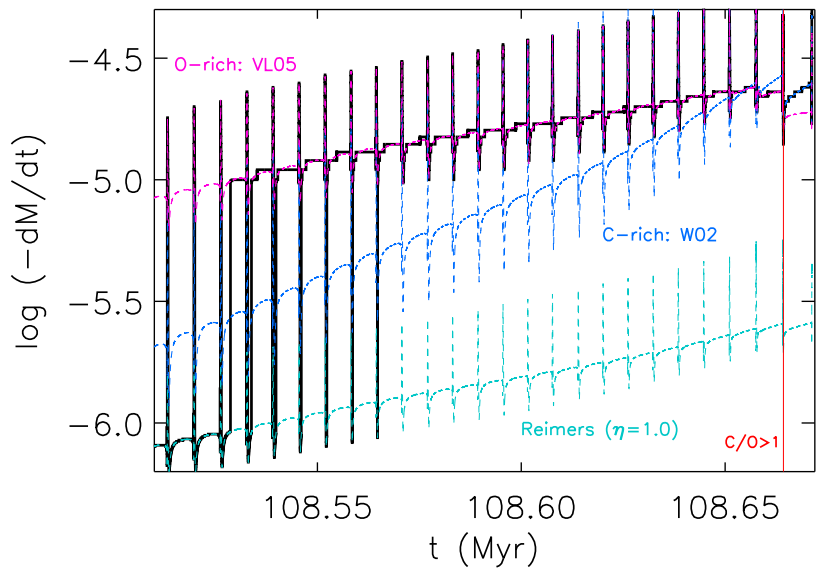}}
\caption{Logarithmic mass-loss rate with respect
  to time (in million years) during the TP-AGB evolution of a
  5M$_{\odot}$ model with solar type composition ($Z=0.02$). The actually 
  adopted  mass loss value is given 
  by the thick black line. The colored dashed lines give the rates
  from each of the individual
  prescriptions according to Eq.~\ref{e:reim} (Reimers),
  \ref{e:wachter} (W02),  or \ref{e:loon} (VL05). The vertical red
  line marks the beginning of the carbon-rich phase.}
\label{f:3.4}
\end{figure}

Figure~\ref{f:3.4} shows an illustrative example about the relative
size of the mass loss due to the various prescriptions and the rate
actually used in the evolution of a $5\msun$ star of solar-like
composition on the AGB. On the early AGB a Reimers wind of
$\eta_\mathrm{R} = 1$ is assumed; after three 
of the thermal pulses shown in this figure, the critical 400~d
pulsation period is reached and mass loss according to Eq.~\ref{e:loon}
is applied. This leads to an increase in $\dot{M}$ by an order of
magnitude. Before the last TP of the figure dredge-up turns the model
into a carbon star and we switch to the dust-driven wind of
Eq.~\ref{e:wachter} (W02), which is slightly above the van~Loon mass loss
rate (VL05). Notice that both types of wind show a drop at this point,
due to an increase in effective temperature which is caused by the
  carbon-enhanced opacities. This is one of the events we encountered, where a
value of $\mathrm{C/O} > 1$ does not lead to the expected decrease in
$\teff$, and which will be discussed below.  Note also the steeper
rise of $\dot{M}$ with time, i.e.\ luminosity, for the dust-driven wind (W02).

\newpage

\section{Results of the calculations}
\label{s:calculations}

\subsection{Overview about pre-AGB evolution}
\label{s:overview}

Although the evolution up to the TP-AGB phase is of less interest in
the context of this paper, we briefly discuss these initial
phases. Tables~\ref{ta:1} and \ref{ta:2} list lifetimes on the
main-sequence, the RGB (to be understood as the time between the end
of central hydrogen and the beginning of central helium burning), the
core helium burning, and until the onset of the first thermal
pulse. In addition, total mass and core mass at the first TP are included.

In Fig.~\ref{fa:1} we compare main-sequence lifetimes of our models
with those by \citet[K03]{karakas:2003} and \citet[VW93]{vasswood:93}
for which chemical compositions agree. Both sets include metallicities
of 0.004 and 0.008, K03 also has a case with $Z=0.02$, but VW93 provides
$Z=0.016$ only. The latter set has always $Y=0.25$, while K03 has
varying helium contents of 0.2476, 0.2551, 0.2928 for the three
metallicities (from low to high $Z$). In addition, both 
sets exclude core overshooting, and have older opacities and nuclear
reaction rates. At the lowest metallicity, where composition
differences are smallest, main-sequence ages between all three sets
agree to better than 10\%. The generally longer lifetimes of our models
with respect to K03, most pronounced for $Z=0.02$, can be understood as a
result of overshooting. Indeed, when repeating the calculations
without overshooting (black triangles in the lowest left panel of
Fig.~\ref{fa:1}) both sets agree to a few per cent. The generally
longer lifetimes of VW93 models can be traced back to their higher
hydrogen abundance, which not only provides more nuclear fuel, but
also decreases surface temperatures due to higher opacities and, because of
the lowered molecular weight, lowers the luminosities of the models.

For the core helium burning lifetimes the discrepancies are much
larger. Our models for stars above $2\,\msun$ have lifetimes shorter
by several factors of 10\%. Similar discrepancies exist between the K03 and VW93
models. Our lifetimes agree best with those of the Padova library
\citep{gbbc:2000}, which we also consulted for comparison. As already
discussed by K03, this phase is very 
uncertain due to its dependence on the
$\Iso{12}{C}(\alpha,\gamma)\Iso{16}{O}$ reaction rate and the treatment 
of overshooting and semi-convection. 

The ignition of helium could be followed in all our models,
independent of whether this happened under non-degenerate conditions
in the intermediate-mass stars, or as a core helium flash in low-mass
stars. No artificial setup of post-flash models was therefore needed
\citep{saw:2005}. 

The pre-AGB phase ends with the first thermal pulse, defined as the
first pulse in which a helium luminosity of $10^4\,L_\odot$ is
reached. This definition coincides either 
  with the first appearance of the
  pulse-driven convection zone, or with one pulse
  earlier than that. In general, such a pulse is not yet fully
  developed.
At this stage, the mass of the hydrogen-free core is a 
quantity of interest, as the future growth of the core is important
for the initial-final-mass relation. In Fig.~\ref{fa:1}, right column,
we compare the core mass at the first thermal pulse with the
values by K03. While they agree very well for the intermediate mass
stars, ours are systematically lower for low-mass stars. In
particular they show the characteristic minimum value around
$2\,\msun$, while the K03 models show only a very shallow variation
below $2.5\,\msun$. In the case of $Z=0.02$ we also added values
obtained by Miller~Bertolami (2008, private communication). They are
shown as blue diamonds and agree very well with our
numbers. Miller~Bertolami uses the same overshooting prescription as
we do. Ignoring overshooting, core masses tend to be smaller for those
stars with convective cores on the main sequence. These core masses
are displayed for the same metallicity as black triangles. In view of
the fact that K03 did not include overshooting, the agreement between 
her core masses and those of our models is in fact another sign of
discrepancy.   

We close this overview by mentioning that the influence of
$\alpha$-enhancement at identical $Z$ is such that lifetimes are
generally shorter and core masses at the first TP slightly larger (see
Tables~\ref{ta:1} and \ref{ta:2}).

\subsection{AGB synopsis}
\label{s:agbsyn}

\subsubsection{Overall properties}
\label{s:agbprop}
\begin{figure}
 \begin{center} 
\includegraphics[scale=0.6]{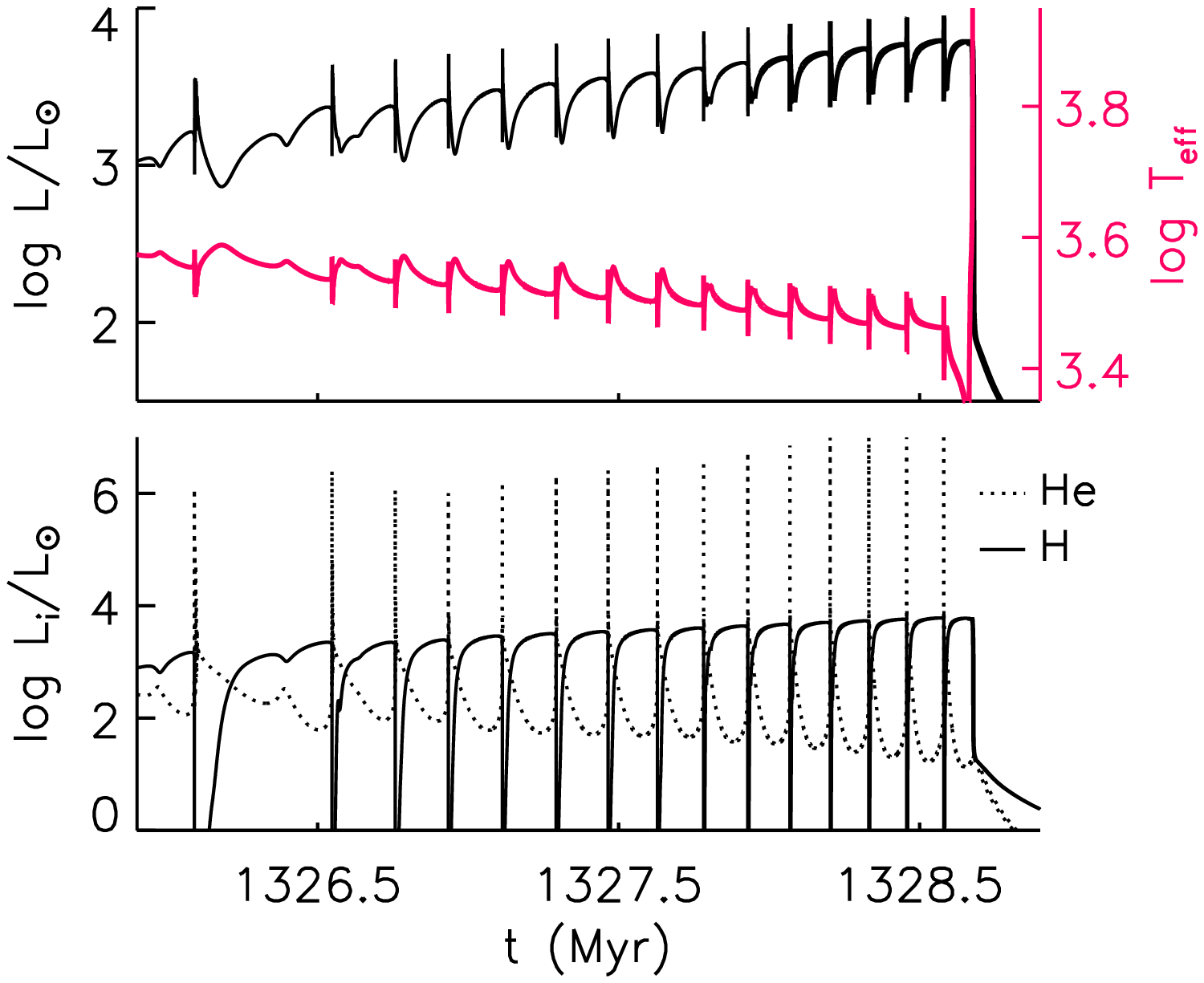} \\  
\includegraphics[scale=0.6]{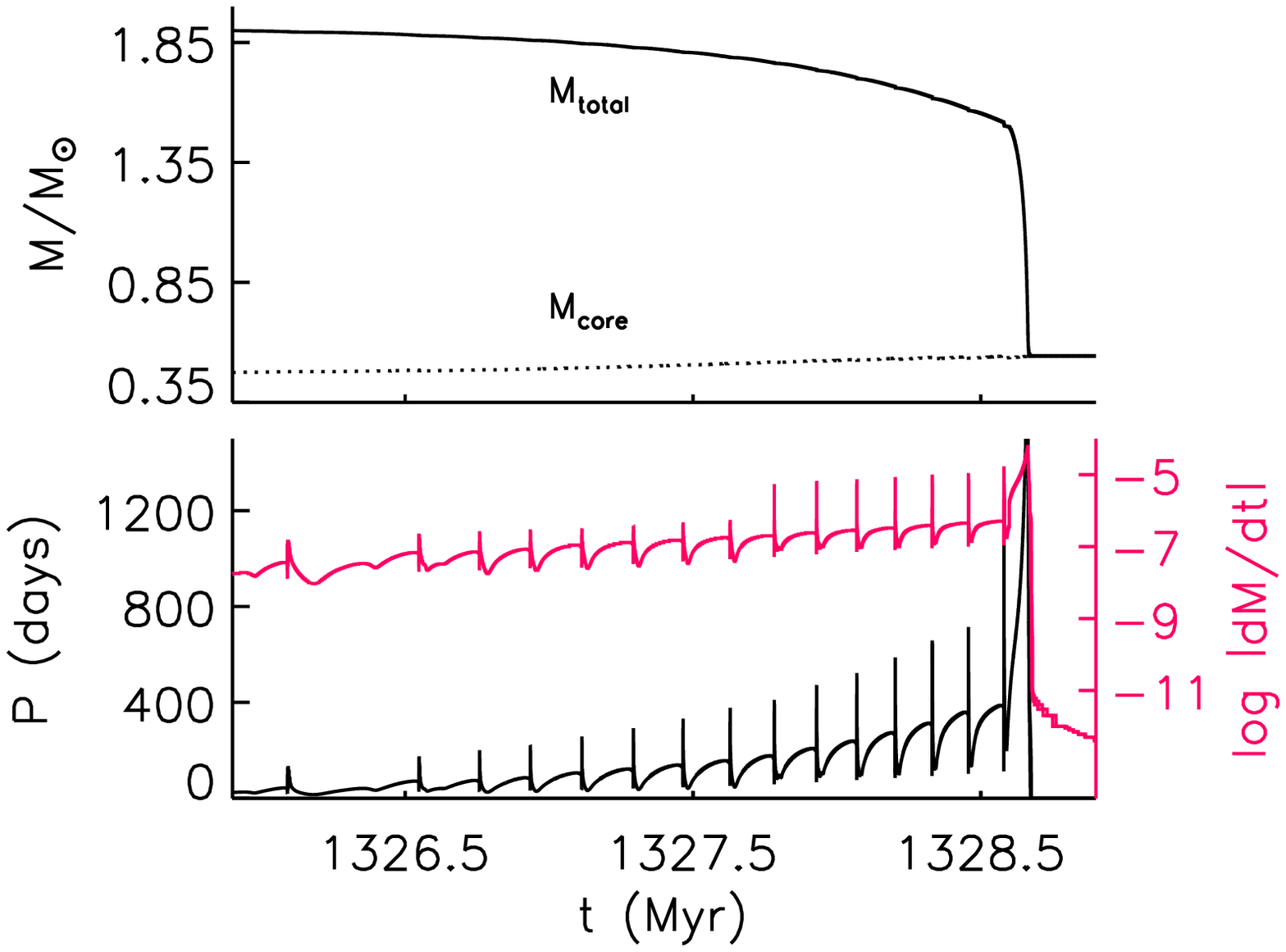} \\
\includegraphics[scale=0.55,bb=51 500 550 710]{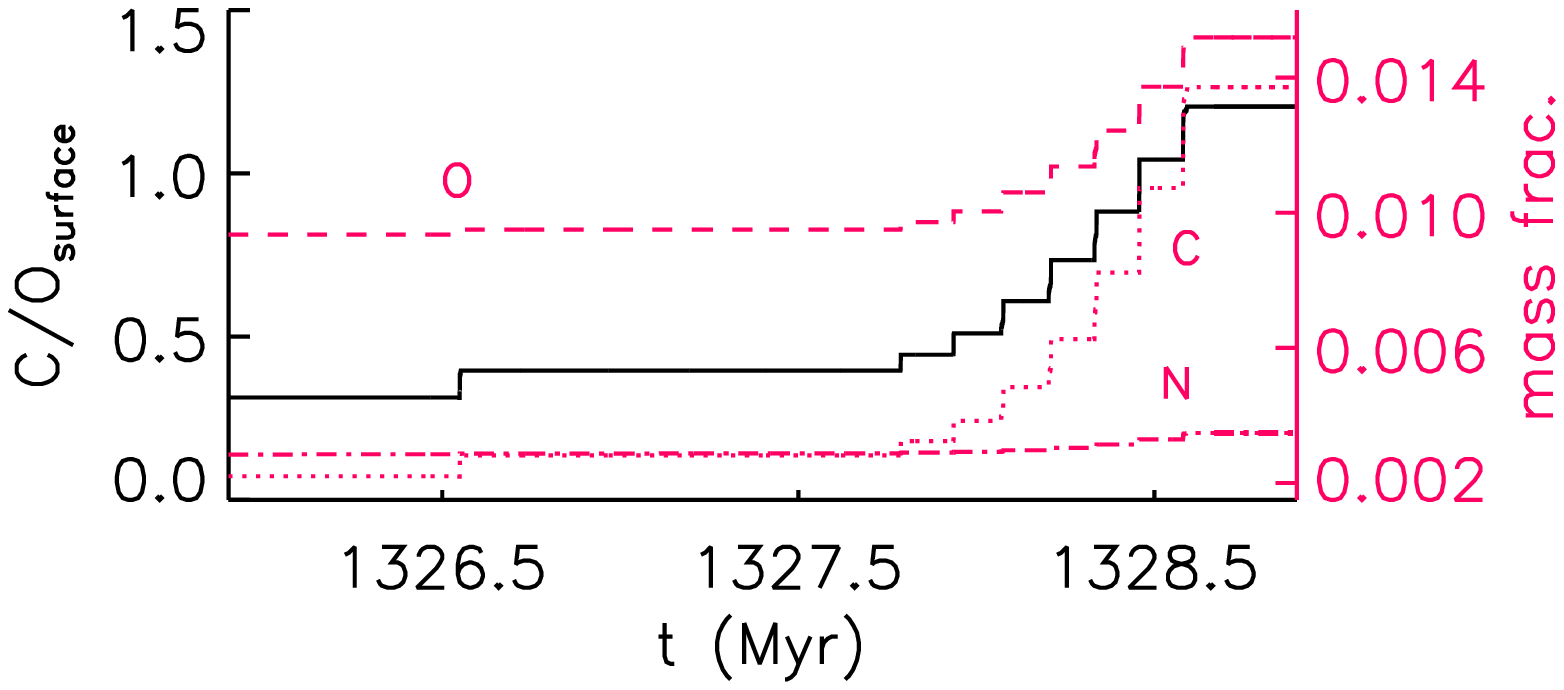}
\caption{Various quantities of a model with initial mass $M=2\,\msun$,
  $Z=0.02$ (solar metal ratios) from the beginning of the AGB 
  until the early post-AGB phase. From top to bottom the panels show
  $\log L$ and $\log\teff$, $\log L_\mathrm{H}$ (solid) and $\log
  L_\mathrm{He}$ (dotted), total and core mass, pulsation period and
  $\log \dot{M}$, C/O-ratio and abundances of C, N, and O. }
\end{center}
\label{f:4.1}
\end{figure}

The evolution of a sample model ($M=2\,\msun$, $Z=0.02$ with solar metal
ratios) along the AGB is shown in Fig.~\ref{f:4.1}. We plot key
quantities such as luminosity, effective temperature, pulsation
period, mass loss rate, and C/O-ratio at the surface.
 The model
experiences in total 15 TPs; the third dredge-up (3du) starts 
at the second TP and continues until the end of the AGB; the mass loss
rate switches from an enhanced Reimers wind to $\dot{M}_\mathrm{AGB}$
according to Eq.~\ref{e:loon} at TP 9, when the pulsation period
reaches the critical 400~days, and finally to the dust-driven
$\dot{M}_\mathrm{AGB}$ (Eq.~\ref{e:wachter}) stage once C/O $> 1$ is reached
during the second last TP. At this epoch a clear drop in $\teff$ is
visible, which is the direct consequence of the carbon-rich opacity
tables we are using. The strong dust-driven wind leads to a
``superwind'' which removes within one interpulse period more than
$1\,\msun$. The remaining, nearly bare core starts the post-AGB
evolution at much higher $\teff$ and a reduced stellar wind. This
model does not experience HBB, as the temperature at the bottom of the
convective envelope reaches temperatures of only around 30~MK, much
less than the typical HBB temperature of 50~MK.

\begin{table*}
\begin{center} 
\caption{Basic model properties along the AGB
  for a total metallicity of $Z=0.02$ and solar and
  $\alpha$-enhanced metal distributions.}
\label{t:4} 
\begin{tabular}{ccccccccccrrr}
\hline
$M_{\mathrm{ZAMS}}^a$ & No & $M_\mathrm{c}$(1)$^c$ & $M_\mathrm{c}$(3du)$^d$
& No & C/O$_{\mathrm{f}}$$^f$ & $M$(f)$^g$ & $M_\mathrm{c}$(f)$^h$
& No & No & \multicolumn{3}{c}{TP-AGB lifetime (10$^{3}$ yrs)$^l$} \\
\cline{11-13}
& TPs$^b$ & & & TP$_{\mathrm{3du}}$(i)$^e$ & & & & TP$_{\mathrm{HBB}}$(i)$^i$ &
TP$_{\mathrm{HBB}}$(f)$^k$ & $t_{\mathrm{mod}}$ & $\approx t_{\mathrm{e}}$
(TPs) & $t_{\mathrm{TP}}$  \\
\hline
\multicolumn{13}{c}{$Z = 0.02$ / solar (mixture III)} \\
\hline
1.0 & 2 & 0.501 & 0.508 & 1 & 0.547 & 0.547 & 0.508 & - & - & {327.675} & {36 (0)} &{364} \\
1.2 & 5 & 0.509 & - & - & 0.361 & 0.608 & 0.524 & - & - & {537.625} & {25 (0)} & {563} \\
1.5 & 8 & 0.511 & 0.526 & 5 & 0.724 & 0.637 & 0.539 & - & - & {911.274} & {15 (0)} & {926} \\
1.6 & 9 & 0.508 & 0.522 & 5 & 0.753 & 0.888 & 0.538 & - & - & {1066.336} & {69 (0)} & {1135} \\
1.8 & 8 & 0.496 & 0.503 & 2 & 1.252 & 1.031 & 0.524 & - & - & {1372.882} & {159 (1)} & {1532} \\
2.0 & 15 & 0.478 & 0.484 & 2 & 1.204 & 0.543 & 0.543 & - & - & {2581.483} & {-} & {2581} \\
2.6 & 14 & 0.518 & 0.533 & 5 & 1.426 & 1.802 & 0.560 & - & - & {1560.967} & {311 (2)} & {1872} \\
3.0 & 12 & 0.596 & 0.596 & 1 & 1.407 & 2.029 & 0.617 & - & - & {693.755} & {141 (2)} & {835} \\
4.0 & 14 & 0.765 & 0.766 & 2 & 0.711 & 2.137 & 0.783 & - & - & {156.353} & {54 (3)} & {210} \\
5.0 & 29 & 0.830 & 0.831 & 2 & 1.051 & 2.354 & 0.850 & - & - & {169.289} & {60 (7)} & {229} \\
6.0 & 34 & 0.928 & 0.928 & 1 & 0.226 & 3.178 & 0.937 & 3 & 29 & {98.985} & {71 (23)} & {170} \\
\hline
\multicolumn{13}{c}{$Z = 0.02$ / $\alpha$-enhanced (mixture IV)} \\
\hline
1.0 & 2 & 0.502 & - & - & 0.164 & 0.551 & 0.505 & - & - & {319.945} & {43 (0)} & {363} \\
1.2 & 5 & 0.508 & 0.518 & 4 & 0.384 & 0.561 & 0.521 & - & - & {529.722} & {12 (0)} & {542} \\
1.5 & 8 & 0.514 & 0.530 & 5 & 1.834 & 0.542 & 0.540 & - & - & {929.208} & {-} & {929} \\
1.6 & 9 & 0.513 & 0.523 & 4 & 0.859 & 0.545 & 0.544 & - & - & {1018.272} & {-} & {1 018} \\
1.8 & 11 & 0.501 & 0.509 & 4 & 1.240 & 0.551 & 0.540 & - & - & {1412.476} & {2 (0)} & {1 414} \\
2.0 & 14 & 0.488 & 0.494 & 2 & 1.248 & 0.629 & 0.537 & - & - & {2196.938} & {9 (0)} & {2 206} \\
2.6 & 15 & 0.513 & 0.527 & 5 & 1.453 & 1.853 & 0.547 & - & - & {1821.007} & {412 (3)} & {2 233} \\
3.0 & 14 & 0.585 & 0.588 & 2 & 1.332 & 1.922 & 0.612 & - & - & {907.181} & {207 (2)} & {1 114} \\
4.0 & 13 & 0.766 & 0.766 & 1 & 0.681 & 1.958 & 0.774 & - & - & {158.834} & {37 (2)} & {196} \\
5.0 & 21 & 0.846 & 0.846 & 2 & 0.541 & 3.215 & 0.860 & - & - & {117.461} & {94 (15)} & {211} \\
6.0 & 37 & 0.946 & 0.946 & 1 & 0.252 & 3.260 & 0.954 & 1 & 32 &
{91.610} & {58 (19)} & {150} \\
\hline \noalign{\smallskip
$^a$initial mass on ZAMS; $^b$number of thermal pulses; $^c$core mass
  at first TP; $^d$mass at onset of third dredge-up; $^e$TP number 
  of 3du onset; $^f$final C/O ratio; $^g$final mass; $^h$final core
  mass; $^i$first TP with HBB; $^k$last TP with HBB; $^l$TP-AGB
  lifetimes (see text). Equivalent tables for all other mixtures are contained in
  the on-line material (Tables~\ref{ta:3}--\ref{ta:6})}
\end{tabular}
\end{center}
\end{table*}

For this metallicity of $Z=0.02$ Table~\ref{t:4} summarizes some key
properties of our models along the AGB. Equivalent tables for the
other four metallicities are contained in the on-line material. 

Despite all numerical improvements, we still encountered convergence
problems such that only part of the models could be evolved to the
very end of the AGB, which we defined as the point when the model
leaves the AGB towards hotter $\teff$ and the pulsation period has
dropped below 100~days. These convergence problems can be traced back
to a dominance of radiation pressure within the convective envelopes,
and have been reported and identified as early as in \citet{wf:86},
but also by, e.g., \citet{karakas:2003}, Herwig (2005, private
communication), \citet{mba:2006}, and
\citet{klpp:2007}. In some cases the convergence problems could be circumvented by
artificially stripping off the remaining envelope and relaxation of
the resulting model, by increasing the mixing length parameter, or by
shifting more mass into an energetically inert outer envelope (which
our models do not have). None of these methods worked satisfactorily
for us, such that we decided to stop the calculation when convergence
problems became insurmountable. In most cases, in particular for lower
masses, the models evolved to virtually the end of the AGB phase. For
example, the $1\,\msun$ star of Table~\ref{t:4} crashed during the
final flash, when it was departing from the AGB during
the third dredge-up. In such cases, due to the amount of mass loss and
the size of the remaining envelope, an estimate of the  TPs and time
still needed to complete the TP-AGB phase can be safely done. This
estimate is given in Table~\ref{t:4} as the second last column
($t_\mathrm{e}$; missing TPs in 
brackets) and added in the final column to the actual AGB lifetime at
which the end of the computations was reached ($t_\mathrm{mod}$). We
follow the approach by \citet{karakas:2003}.
From
Tables~\ref{t:4} and \ref{ta:3}--\ref{ta:6} it is evident, that up to
$M=4\,\msun$ at most 2 TPs are missing for the complete AGB evolution,
but that for higher masses a substantial part of the AGB is
missed. Altogether 100 out of 110 sequences reached that advanced
stage, with 79 missing less than 3 TPs. For the two lowest metallicities
some models experienced 
convergence problems so early on the TP-AGB that we did not include
them in these tables. 

\begin{figure} 
\centerline{\includegraphics[angle=90,scale=0.40]{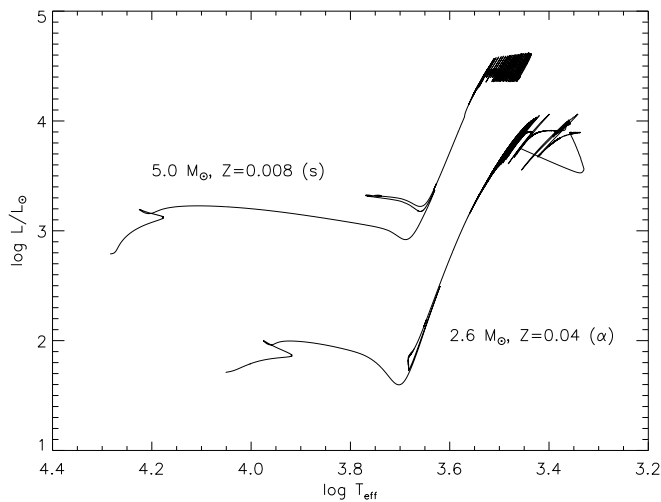}}
\caption{Evolution of two of our models. Shown are stars of $5\,\msun$
with $Z=0.008$ (solar-scaled) and $2.6\,\msun$ with $Z=0.04$
($\alpha$-enhanced).}
\label{f:4.11}
\end{figure}

We show as examples the evolutionary paths of two of our models
in the Hertzsprung-Russell-Diagram (Fig.~\ref{f:4.11}), calculated
with all physical details discussed in Sect.~\ref{s:progagb}. The
$5\,\msun$ star (of the typical LMC composition V) experiences 38 TPs in
total. Dredge-up occurs from the first TP, and is
sufficient to turn the star into a carbon star. The dredge-up is
competing, however, with strong HBB between TPs 9 and 24, which reduces
C/O to a final value close to 1. During the
last three pulses the convergence problems increased and the evolution
in the HRD became irregular. We do not show these last TPs. 

The second star with $M=2.6\,\msun$ and an $\alpha$-enhanced mixture
of $Z=0.04$ (possibly representative for an extreme bulge composition)
experienced 16 TPs and dredge-up after the $9^\mathrm{th}$ TP, which
increased $Z$ to $0.048$. The decrease of $\teff$ between pulses,
clearly visible in Fig.~\ref{f:4.11} is the consequence of this overall
metallicity increase due to dredge-up of mainly C and O, and to a
lesser extent of N. Although C/O more than doubles to 0.432, the star
does not become a carbon star; mass loss is strong according to
Eq.~\ref{e:loon}. This star finally leaves the AGB during the last TP,
and turns to hotter temperatures, but the calculations ended before it
reached the post-AGB phase, due to the mentioned convergence problems.

\subsubsection{Influence of composition}
\label{s:compinf}
Our models recover some previously known trends with mass and
metallicity. The 3du is more pronounced for lower metallicity and
higher mass. This is also reflected in the C/O-values. For the highest
masses, HBB reduces carbon. It sets in earlier for lower
metallicity. While this is consistent with, e.g.,
\citet{karakas:2003}, the C/O ratios reached in her models are
generally higher than ours, up to 20 compared to about 5 in our
case. It should be kept in mind that the amount of dredge-up depends
crucially on the method to obtain it: while we use overshooting,
\citet{karakas:2003} achieves it by extending the convective regions to
a point of marginal stability, but does so only for the lower
  boundary of the convective envelope. Her models do not contain any
  overshooting from the pulse-driven convective layers, and therefore
  less C and O are mixed from the core into the intershell layers. As
  \citet[Figs.~11 and 12]{herwig:2000} has demonstrated, overshooting
  leads to relatively more oxygen than carbon enhancement, and therefore
  to lower C/O-ratios. This is a reason for our lower
  surface C/O-ratios. 
According to \citet{herwig:2000} and 
  \citet{wh:2006} the present surfaces of hot, hydrogen-deficient
  post-AGB stars of type PG1159 and [WC] have formerly been interior
  layers of AGB stars, uncovered by late AGB or post-AGB thermal
  pulses. If that is indeed the case, these stars present a valuable
  test for the intershell composition of stars along the AGB. In most
  cases \citep[see Table~1 of][]{wh:2006}, the observed C/O-ratio is below 10,
  and clusters around values of 3--5, which would confirm the lower
  values resulting from the inclusion of overshooting from the
  He-driven convective zone. Similarly, C/O-ratios in planetary
  nebulae are, in spite of all difficulties in determining them,
  in most cases well below our maximum value of 5 \citep[see][for a
    comprehensive and thorough analysis]{liuliu:2004}, and never reach
  values close to 10 \citep[see also][for two PNe in the Sagittarius
    dwarf galaxy]{pzwd:2000}.

HBB occurs in the models of \citet{karakas:2003} and in those
by \citet{herwig:2004b} around $M=3.5\,\msun$; our models show
this only at higher masses ($M=5\,\msun$ and above). 

A new feature of our grid is the use of solar-scaled and
$\alpha$-enhanced metal distributions. We find that there is no
significant influence on the number of TPs or the occurrence of 3du and
HBB. The core mass at the onset of TPs and the 3du tends to be larger
by a few $10^{-3}\,\msun$ and the final C/O ratio tends to be
lower. Since C/O is initially only 0.19 in the
$\alpha$-enhanced mixtures, it is clear that an enhanced 3du is needed to
convert the model into a carbon star. As a general rule, the
$\alpha$-enhanced models need about 1-2 TPs more to achieve a similar
C/O-ratio. This is the reason why the $2.6\,\msun$ model of
Fig.~\ref{f:4.11} does not turn into a carbon star in spite of dredge-up.

\begin{figure}
\begin{center} 
\includegraphics[scale=0.5,angle=90]{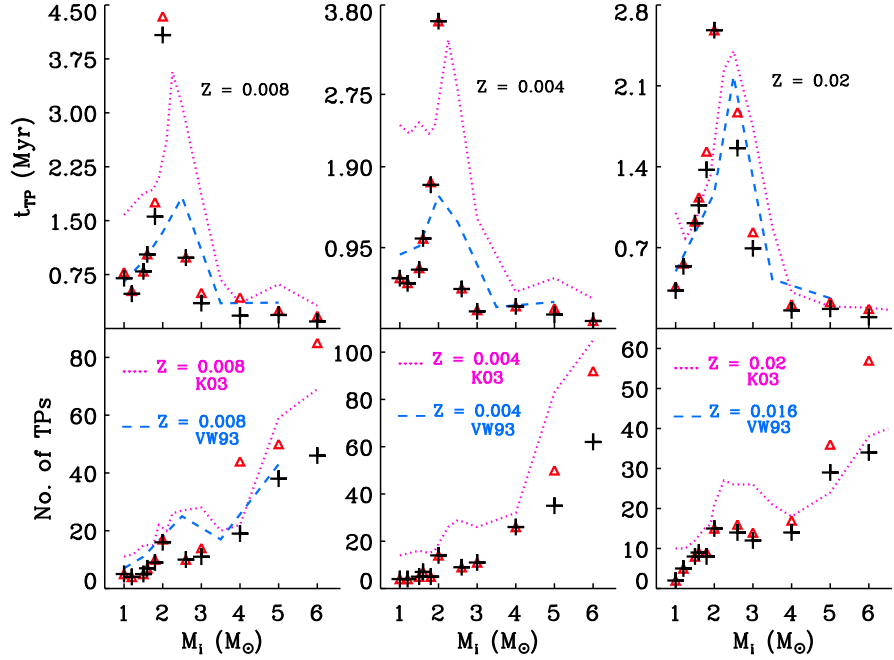}
\caption{Top panels: Lifetimes on the TP-AGB for our models (black
  crosses) in comparison with those by K03 (red dotted lines) and VW93
(blue dashed lines) for the three metallicities in common. The most
  metal-rich mixture of VW93 is $Z=0.016$ instead of 0.02,
  however. The red triangles are our estimates for the total lifetimes
  $t_{\mathrm{TP}}$ of our models taking into account $t_{\mathrm{e}}$ of
  Table~\ref{t:4}. 
Bottom panels: comparison of the number of thermal pulses. Symbols and
lines as above. Data for the two more metal-rich mixtures of VW93 are
not available.}
\label{f:4.2}
\end{center}
\end{figure}

\subsubsection{Comparison with other models}
\label{s:agbcomp}
\begin{figure}
\begin{center} 
\includegraphics[angle=90,scale=0.4]{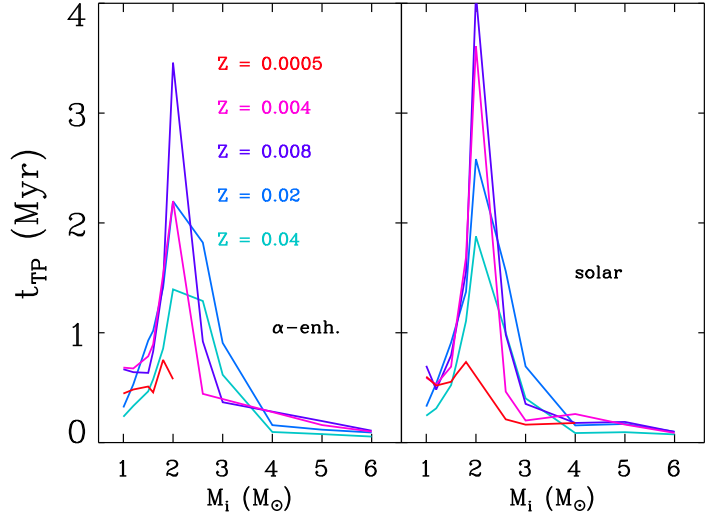}
\caption{Lifetimes on the thermally pulsing AGB as function of initial
  mass for all our models and
chemical compositions. Left: $\alpha$-enhanced metal ratios; right:
solar-scaled.}
\end{center}
\label{f:4.3}
\end{figure}

The comparison with the results of \citet{karakas:2003} and
\citet{vasswood:93} with respect to TP-AGB lifetimes and number of
thermal pulses is shown in Fig.~\ref{f:4.2}. Our models show the same
global behaviour with a maximum of $t_{\mathrm{TP}}$ at the lowest
initial mass that starts core helium burning under non-degenerate
conditions and therefore has the lowest core mass, which leads to the
longest interpulse times. Note, however, that the K03 models for
$Z=0.004$ show an untypical behaviour at the lowest $M_i$. Lifetimes
and number of thermal pulses of our models are typically lower than in
the other two calculations, in particular for higher masses. This is
due to our mass loss formulas, which lead to globally higher mass loss
rates, in particular after C/O-ratio exceeds unity
(Fig.~\ref{f:3.4}). Although the differences in the physical input for
the calculations are larger in comparison with VW93, Fig.~\ref{f:4.2}
shows a better agreement with these older calculations. The larger
differences between K03 and VW93 (both use the same mass loss
description) have no explanation. The influence of the metal
distributions (solar-scaled or $\alpha$-enhanced) is shown in
Fig.~\ref{f:4.3}. Generally, lifetimes on the TP-AGB are shorter for
the $\alpha$-enhanced mixtures, although the number of pulses tends to
be slightly higher (see above). 

\subsubsection{Final core mass}
\label{s:ifmr}
\begin{figure}
\begin{center} 
\includegraphics[angle=90,scale=0.4]{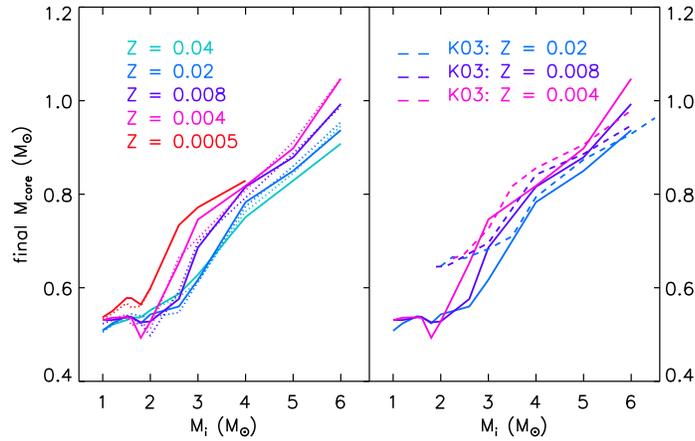}
\caption{Left: Final core mass of all our models (solid lines: solar-scaled
  metal ratios; dotted lines: $\alpha$-enhanced). Right: Comparison
  with K03 results (dashed).}
\end{center}
\label{f:4.4}
\end{figure}

As another global quantity we present in this section
the final core mass, shown in Fig.~\ref{f:4.4}, at the end of the TP-AGB
evolution for all ten mixtures as well as the comparison with the
results by K03 for the three mixtures in common. The comparison shows
a generally good agreement with K03 for higher stellar masses, but much lower
$M_\mathrm{core}$ values at lower stellar masses. This is the consequence of
the lower core mass at the first TP, visible also in Fig.~\ref{fa:1},
and the effect of our overshooting prescription, which prevents the
core from growing substantially. This indicates that our overshooting
prescription, in particular always using the same overshooting
parameter at all convective boundaries may lead to an overestimate of
the effect of overshooting. \citet{sswm:2009} and \citet{lhlgs:2003}
have argued that the overshooting from the base of the
pulse-driven convective layer in the He-shell should be somewhat 
smaller during the core hydrogen burning to allow the core
to grow and to be in better agreement with 3d-hydrodynamical simulations by
\citet{herwig:2007}. In terms of the parameter $f$ in Eq.~\ref{e:ov}
the numerical value should be $f \lesssim 0.01$ instead of our standard value
of 0.016. Indeed, core masses of our models grow during the TP-AGB
phase by less than $0.02\,\msun$ except in the mass range between 2 and
$4\,\msun$, where the growth reaches $0.04 \cdots 0.06\,\msun$ (more
for larger metallicities). 

\begin{figure}
\begin{center} 
\includegraphics[angle=90,scale=0.45]{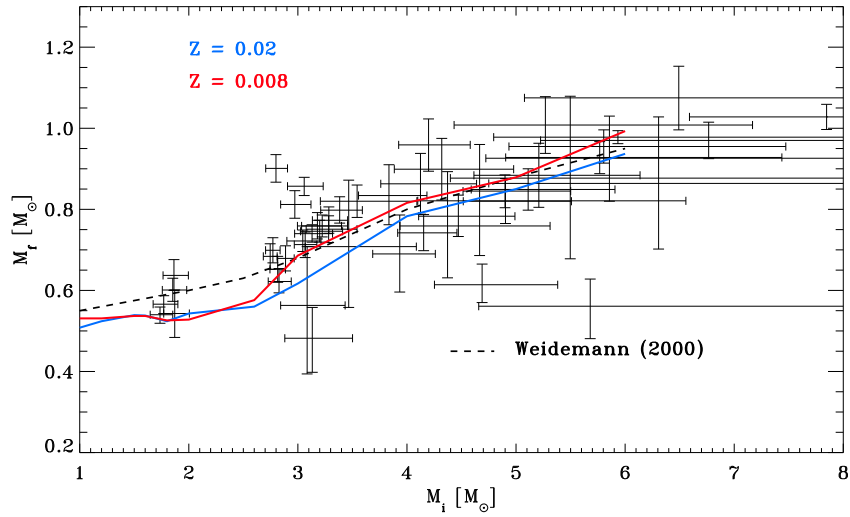}
\caption{Our predicted initial--final--mass--relation (solid lines) for
  $Z=0.02$ and $0.008$ (solar metal ratios) in comparison with
  observed open cluster objects and the empirical fit by \citet[dashed
    line]{weide:00}. 
The initial ($M_\mathrm{i}$) and final ($M_\mathrm{f}$) mass values
and associated error bars are taken from \citet{sswm:2009}.}  
\end{center}
\label{f:4.5}
\end{figure}

Our predicted initial--final--mass--relation (IFMR) is therefore very
close to the relation between initial mass and core mass at the first
TP. This is evident from Fig.~\ref{f:4.5}, which shows that our IFMR
is a lower envelope -- at least at the low-mass end --  to the
empirical data for clusters with 
$\mathrm{[Fe/H]} \approx 0.0$. The initial solar metallicity lies
somewhere between the two cases shown. This constitutes a mild
discrepancy with the observations for the lower initial mass range,
where our predicted IFMR drops below the semi-empirical relation by
\citet{weide:00}. It agrees well with one by Miller~Bertolami (2007; private
communication), shown in \citet{sswm:2009}, who used a very
similar overshooting description as we do. The final core masses of
K03, shown in Fig.~\ref{f:4.4}, on the other hand, are higher than
the empirical relation for $M \leq 3\,\msun$.

\subsubsection{Analytical fits}
\label{s:anfit}
Synthetic AGB calculations make use of analytical relationships
between key quantities of full AGB models. Among them, the core-mass
-- luminosity and the core-mass -- interpulse-period relations are two
crucial ones. A number of such relations are available in the
literature. We mention here the early linear one by
\citet{paczynski:70}, the one used by \citet[M96]{mbc:96}, which includes a
dependency on the envelope composition, and the two most recent and
complex relations by \citet[WG98]{waggroen:98} and
\citet[I04]{itkp:2004}. They are based on the models by
\citet{wag:96} and \citet{karakas:2003}, respectively, and 
also include dependencies on composition, mixing length parameter, hot
bottom burning and dredge-up. Since 
the latter two effects differ between calculations, it
cannot be expected that such detailed relations agree very well with
our new models, which incorporate the additional effects of
overshooting, opacities for varying C/O-ratios, and mass loss
prescription.  Nevertheless, a
detailed comparison was done by \citet{kitsikis:2008}, which we will
not repeat here. He found that with respect to the  core-mass --
luminosity relation, the one by M96 agrees best for low-mass
models, because our models have generally lower core masses for a given
luminosity. This is related to the low-mass discrepancy found above
for the IFMR. For $M_\mathrm{core}\gtrsim 0.7\,\msun$ both the I04 and WG98
relations describe our models equally well. The deviations, however,
always remain within the 10 to 20\% range. 
With respect to the core-mass -- interpulse-period relation the WG98
description fits our models best. Part of the deviation were
traced back by \citet{kitsikis:2008} to the effect of the new
opacities on luminosity and interpulse time. A new adaption of the
WG98 analytical description to our new models seems promising, but
awaits completion. If synthetic models require an average accuracy at
the 10\% level, the fits by WG98 and I04 might still be used.

\subsection{Effect of varying C/O-ratio}

\begin{figure}
\begin{center} 
\includegraphics[scale=0.45,angle=90]{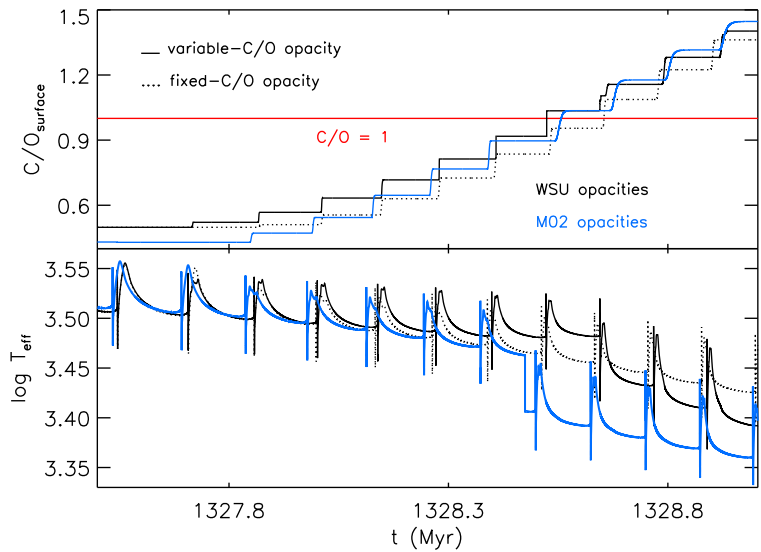}
\caption{Evolution of a model with $M=2\,\msun$ and $Z=0.02$ (solar)
  on the AGB using different sets of opacity tables. Black lines
  correspond to the new WSU molecular opacities (Sect.~\ref{s:wichita}),
  the blue one to the approximative molecular opacities by
  \citet[Sect.~\ref{s:marigo}]{marigo:2002}. The black dotted 
  line refers to calculations where only the total metallicity, but not
  the change in C/O due to the third dredge-up is accounted for.}
\end{center}
\label{f:4.6}
\end{figure}

Since we emphasized a consistent treatment of the carbon enrichment of
the envelopes in this work, we will now discuss in detail how this
influences the models. We start with a comparison of cases with
different sets of opacity tables. This illustrative test was done for
a solar composition (mixture III)
and a mass of $2\,\msun$. Mass loss was strongly reduced to
prevent an amplification of differences in the models due to sensitive
mass loss rates. The C/O-ratio and the effective temperature evolution
are shown in Fig.~\ref{f:4.6} for three different cases. The early AGB
evolution up to the 7th TP is not shown. Different 3du histories
explain why the C/O ratio is not the same in all cases.

The black dotted line corresponds to the use of opacity tables with no
specific treatment of the C/O-ratio. This implies that any increase in
carbon abundance is taken into account only by using opacity tables of
higher metallicity, but still solar metal ratios. The solid blue line
is using the \citet{marigo:2002} molecular opacities. Here, any
increase in metallicity is ascribed to be due to C only, irrespective
of the actual C/O ratio. Generally, the C/O ratio is overestimated in
these tables, such that the influence on the opacities is exaggerated.
These molecular opacities were taken into account in our
  calculations, when the C/O ratio in the models exceeded unity, or
  when the total metallicity reached $Z\approx 0.03$, which is the
  case shown in the figure. 
This implies a sudden jump in opacity, and a corresponding
sudden decrease in $\teff$, visible at $t=1328.5$~Myr. As such a jump
creates in some cases numerical problems in the calculations, we
applied the switch to the molecular opacities during the interpulse
phase. The lower $\teff$ then leads to stronger dredge-up in the
following TPs. 
Compared to the
equivalent case of ignoring carbon enhancement in the opacities,
$\teff$ drops by up to 0.07~dex during the final 5 TPs. The level of
carbon enhancement (upper panel of Fig.~\ref{f:4.6}) is increased with
respect to the fixed C/O case.

The WSU opacity tables, introduced in Sect.~\ref{s:wichita}, in contrast
follow the C/O-ratio in detail, therefore the slow increase in carbon has
a smooth influence on $\teff$ (black solid line).While the WSU molecular
opacities also lead to a decrease in $\teff$, interestingly initially,
when C/O is approaching unity, this decrease of $\teff$ is less pronounced
than in the case where carbon enhancement is taken into account as a
general increase of an otherwise solar-scaled metallicity (dotted
black line). However, when looking at Fig.~\ref{f:3.2}, it is evident that
indeed up to $\log\teff \approx 3.5$ the Rosseland mean opacity first {\em
  decreases} with increasing C/O and only for $\mathrm{C/O} > 1$ begins to
increase again. Although the physical conditions ($Z$, $\log R$) of that
figure are not quite the one encountered in the models shown in
Fig.~\ref{f:4.6}, this case is very representative. The reason for the
initial drop is the reduced number of TiO and H$_2$O molecules, which for
O-rich mixtures are a major source of opacity. Eventually, when
$\mathrm{C/O} \gtrsim 1.2$, the C/O-variable opacities are higher and
$\teff$ drops stronger. In our set of tables of
Marigo's molecular opacities, we do not have mixtures with
$\mathrm{C/O}\approx 1$, such that the opacity minimum cannot be
recovered. This explains why the solid blue line does not show this
weaker decrease of the effective temperature as a consequence of carbon
enhancement. We emphasize that $\teff$ in all cases continues to decrease
during the TP-AGB phase, but at different rates depending on the detailed
treatment of C-enhancement.

\begin{figure} 
\begin{center}
\includegraphics[scale=0.45,angle=90]{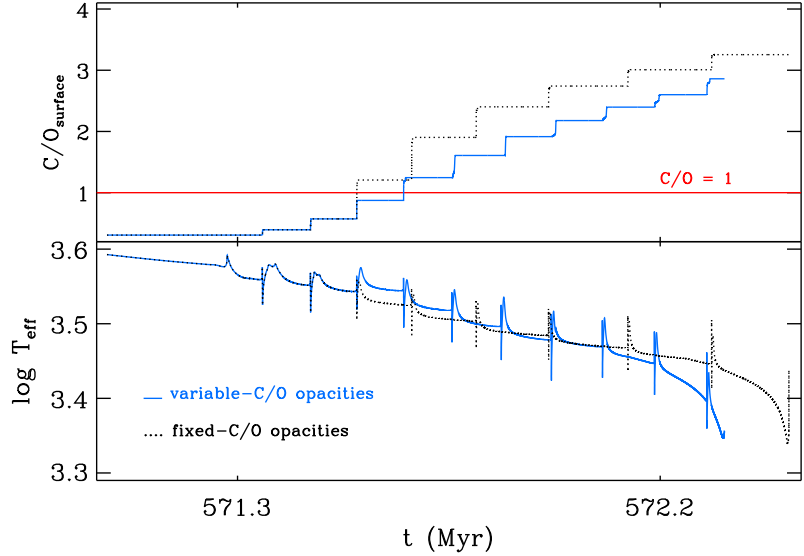}
\caption{Evolution of a model with $M=2.6\,\msun$ and $Z=0.008$ (solar)
  on the AGB using the new WSU molecular opacities. Variable
  C/O-ratios are taken into account in the case indicated by the blue
  line, and ignored in that shown in black.}
\end{center}
\label{f:4.7}
\end{figure}

A similar case is shown in Fig.~\ref{f:4.7}. Here, we employ our
standard mass loss treatment. We compare only the case of ignoring or
including variable C/O-ratios. The opacity tables (WSU molecular
opacities) are the same in both
cases. When dredge-up begins (at the third full TP), the C/O-variable
opacities again lead to higher $\teff$ as compared to the fixed-C/O
opacity tables. As a consequence, dredge-up is
{\em more effective} for the case, in which the increase in carbon is
only taken into account by using opacities for higher $Z$, but still
solar C/O. Only when $\mathrm{C/O}\gtrsim 2$ (TP no.~6), the increase
in carbon is reflected by higher opacities, lower $\teff$, and a more
efficient 3du, such that the blue line is catching up in the upper
panel of Fig.~\ref{f:4.7}. Although the final $\log\teff$ is about 0.1~dex
lower than in the other case, C/O at the surface is still slightly
lower. As a consequence of the lower $\teff$, mass loss is higher and
the evolution stops earlier in the ``variable opacities'' case.

\begin{figure}
\centerline{\includegraphics[scale=0.45,angle=90]{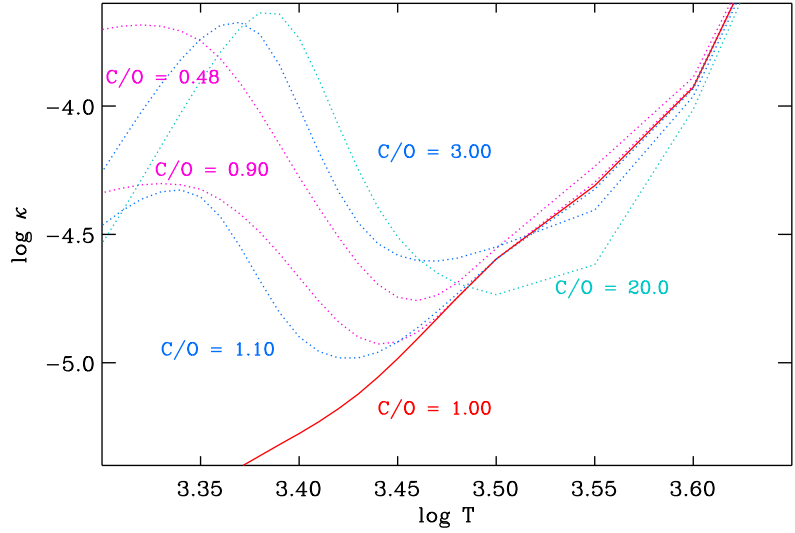}}
\caption{As Fig.~\ref{f:3.2}, but for $Z=0.0003$. The increase in
  $\kappa$ due to strong carbon enhancement has shifted to lower
  temperature. }
\label{f:4.8}
\end{figure}

This effect of an initially shallower $\teff$-decrease as a consequence of
carbon-enrichment of the envelope is more pronounced for lower overall
metallicity, as can be inferred already from Figs.~\ref{f:4.6} and
\ref{f:4.7}. We have verified this with a further test case at our
lowest metallicity ($M=1.8\,\msun$, $Z=0.0005$). The reason is that
with decreasing metallicity the opacity maximum for $\mathrm{C/O} >1$
is shifting from $\log T \approx 3.4$ for $Z=0.02$ (Fig.~\ref{f:3.2})
to $\log T \approx 3.35$ for $Z=0.0003$ (which is the table
metallicity closest to that of the model, see Fig.~\ref{f:4.8}), and
that the C-rich opacities are no longer higher at $\log T \approx 3.5$.
This effect is confirmed by \citet[their Fig.~4]{csla:2007} for their
$2\,\msun$, $Z=0.0001$ model, calculated also with carbon-enhanced
molecular opacities. As a  consequence of the higher temperatures of
C-enhanced models, mass loss is decreased and TP-AGB lifetimes
prolonged. This figure also shows that when C-enhancement is not taken
into account at all, i.e.\ if $Z$ is kept constant, $\teff$ is hardly
decreasing at all. The same influence of an increasing C/O-ratio on the
Rosseland mean opacity can be found in the data by \citet[their
  Fig.~2]{la:2009}.

\subsection{Carbon-stars lifetimes} 

\begin{figure} 
\begin{center}
\includegraphics[scale=0.40,angle=90]{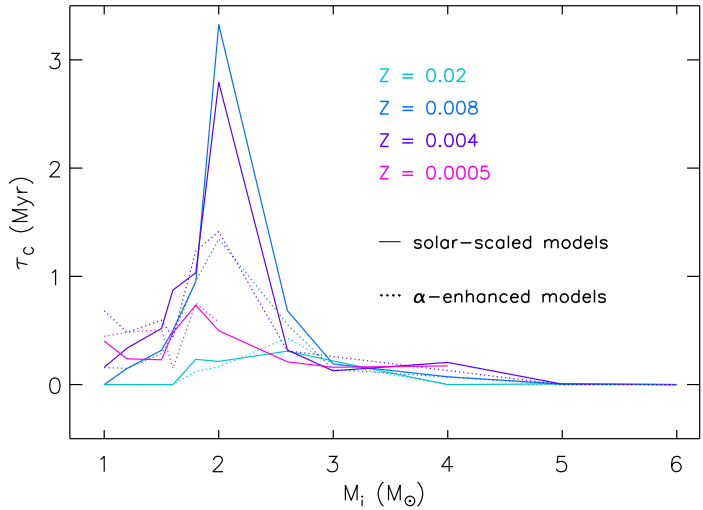}
\caption{Lifetimes of all our models as carbon stars. solar-scaled
  metal mixtures refer to the solid lines, $\alpha$-enhanced ones to
  to the dotted lines.}
\end{center}
\label{f:4.9}
\end{figure}

The interaction between dredge-up, low-temperature opacities, and mass
loss determines for what duration a stellar model is a C-star. Since
our calculations differ from previous full AGB models in those
aspects, we expect that carbon-star lifetimes will be modified
considerably. We show in Fig.~\ref{f:4.9} the C-star lifetimes of our
models for all 10 compositions. The efficiency of the 3du, which rises
with lower metallicity, leads to the pronounced peaks for $Z=0.004$
and $Z=0.008$, but the lowest metallicity models have a much smaller
peak. The location of the lifetime peak shifts to lower initial mass
with decreasing metallicity, although for $Z=0.02$ it is not very
pronounced. Note that the $Z=0.04$ case is missing as for this
metallicity our models do not turn into carbon stars
(Table~\ref{ta:3}). 

\begin{figure}
\begin{center}
\includegraphics[scale=0.45,angle=90]{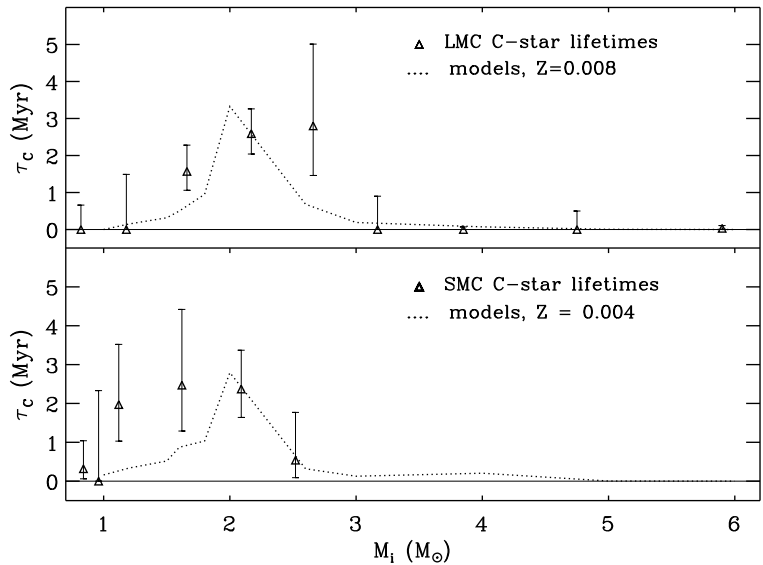}
\caption{Lifetimes of carbon stars, $\tau_\mathrm{C}$ in the LMC
  (upper panel) and SMC 
  (lower panel). The observational data and error-bars are from the
  compilation by \citet{gima:2007}, the dotted lines correspond to our
  model predictions for the two appropriate compositions with
  solar-scaled metal ratios, as given in the figure.}
\end{center}
\label{f:4.10}
\end{figure}

\citet{gima:2007} have derived the C-star lifetime from star counts in
clusters of both Magellanic clouds. Their Fig.~3 shows a comparison of
these with {\em synthetic} AGB population predictions. They note that the
carbon-star luminosity function could be reproduced successfully only
since the work of \citet{gdj:93}, but that even this model, as well as
all previous ones underestimated the C-star lifetime peak for initial
masses below $3\,\msun$, and overestimated it for higher masses. 
\citet{marigo:2002}, introducing the C-variable molecular opacities
managed to obtain an acceptable fit to the observed data. In
Fig.~\ref{f:4.10} we compare the data by \citet{gima:2007} for the LMC
with the predicted lifetimes of our own models with $Z=0.008$ (upper
panel) and for the SMC with models with $Z=0.004$ (lower panel; both
for solar-scaled
metallicities). In both cases, but in particular for the LMC, the
agreement is quite satisfying, although the peak width is too low for
the SMC. Note also that the C-star lifetimes are much lower for
initially $\alpha$-enhanced composition (Fig.~\ref{f:4.9}, dotted
lines). As we mentioned before this is simply a consequence of the
initially lower C/O-ratio that requires more 3du episodes. 

\subsection{Post-AGB evolution}
\label{s:postagb}

At the end of the AGB-phase, the models evolve to higher $\teff$ with
decreasing pulsation period. To date the largest set
of post-AGB tracks are still the 27 tracks  by \citet{vasswood:94} for
the same metallicities as in \citet{vasswood:93}, and covering the
post-AGB mass range from $0.558\,\msun$ to $0.943\,\msun$. 
These post-AGB models are in fact the
continuation of the AGB-models of \citet{vasswood:93}, and constitute
the very rare case of 
continuous evolutionary models that evolve from the main-sequence to
the white dwarf cooling stage. In our case, as mentioned before, we
encountered severe convergence problems for many models, which are
similar to those reported, e.g., by \citet{wf:86},
\citet{herwig:2005}, \citet{mba:2006}, and \citet{klpp:2007}. From the
100 tracks evolved to the end of the AGB, or sufficiently close to it,
only 60 could be followed through the post-AGB phase. They are all from
initial masses in the range of 1 to $2\,\msun$. Half of them could be
computed continuously, for the other half we had to ``freeze'' the
models at the end of the AGB, remove the remaining envelope, and
resume the full evolution at sufficiently higher temperature. In many
cases this was successful only when $\teff \approx 10^4$~K, which is
a standard definition for the start of the post-AGB evolution
\citep{vasswood:94,mgwgc:2004}. The time between the beginning
departure from the AGB and this point is taken as the transition time
$t_\mathrm{tr}$. We derive $t_\mathrm{tr}$ in those cases where we
have continuous models as the time between the model for which the
pulsation period has dropped to 100~days and the $\teff = 10^4$~K
point. As a consequence of the restricted initial mass range for which
we were able to follow the post-AGB evolution, post-AGB stellar masses
ranged only between 0.5 and $0.6\,\msun$. About one third of our models
experience a late or very late TP (LTP or VLTP) during their horizontal
crossing of the HRD respectively on the WD cooling track. The calculations were
stopped when they had returned to the AGB for the second time. Such
models had to be followed in each case separately to overcome the 
numerical difficulties. 

Since our post-AGB grid is restricted in mass, we refer to
\citet{kitsikis:2008} for the detailed results. Here we summarize only
a few key points. Of the 30 cases for which we could follow the
complete evolution without intervention, 4 left the AGB as He-burners
(defined as leaving the AGB through a pulse cycle phase below
0.15). Although this number is as low as that in \citet{vasswood:94}
and \citet{bloecker:95a}, it cannot be compared directly, since in
both these cases most He-burners originate from born-again stars,
i.e.\ stars that experienced a (V)LTP. Including this fact, our fraction
of He-burners would be higher. On the other hand, we can expect that
models with higher initial mass, if we could follow them to the
post-AGB, would reduce the fraction of He-burners. This is because our
mass-loss description favors envelope ejection during the late
TP-cycle only for low-mass stars, while for higher masses the mass
loss rate remains high through a large fraction of the interpulse
phase, too.

The transition times greatly depend upon the definition of the post-AGB
phase and the termination of the final
AGB-superwind. \citet{vasswood:94} begin the post-AGB evolution
between temperatures of 3500~K and 5000~K, while those by
\citet{bloecker:95a} start between 6000~K and 7900~K. Our definition
of the 100~d pulsation period leads to an earlier post-AGB start,
corresponding to $\teff \approx 3800 \cdots 5400$~K. Similarly, the
post-AGB mass loss description influences the transition
speed. \citet{vasswood:94} rather abruptly switch to radiation-driven
winds, which for $\teff < 10000$, the end of the transition phase, is
lower than our adopted Reimers-type wind (Eq.~\ref{e:reim}), which
is similar to the approach by \citet{bloecker:95a}. As a result,
$t_\mathrm{tr}$ ranges from a few hundred to about 2000~yrs for most
of our models, with a slight tendency to increase for lower masses and 
higher metallicities (up to $\approx 10000$~yrs for $Z=0.04$). Two
He-burner post-AGB models, both with $Z=0.04$ and of initial mass
$1\,\msun$ (solar and $\alpha$-enhanced cases) are outstanding and
reach 50000~years. The overlap in post-AGB mass with
\citet{vasswood:94} is very restricted; their $t_\mathrm{tr}$ in the
mass range in common with us is up to ten times longer. According to
\citet{detlef:07} post-AGB stars have $\teff \gtrsim 5000$~K and
transition times around 1000~yrs, as inferred from hydrodynamical
studies. Our models thus show a reasonable agreement with these
results. 

Two further timescales of the post-AGB evolution are of interest. The
first one is the time until the radiation-driven hot wind of
Eq.~\ref{e:cspn} sets in
(technically the point, when it is larger than the Reimers wind of
Eq.~\ref{e:reim}). Since our post-AGB mass loss description is very
similar to that by \citet{bloecker:95b}, we can compare this
timescale $t_1$, but only for the mass in common,
$M=0.6\,\msun$, where his $t_1$ is $1700$~yrs while ours is 
$t_1 \approx 500$~yrs. This is close to that by
\citet{bloecker:95b} for $M=0.625\,\msun$, who finds an increase in
$t_1$ with decreasing mass, strongest for the lowest masses, which we
confirm.

\begin{figure}
\begin{center} 
\includegraphics[scale=0.45,angle=90]{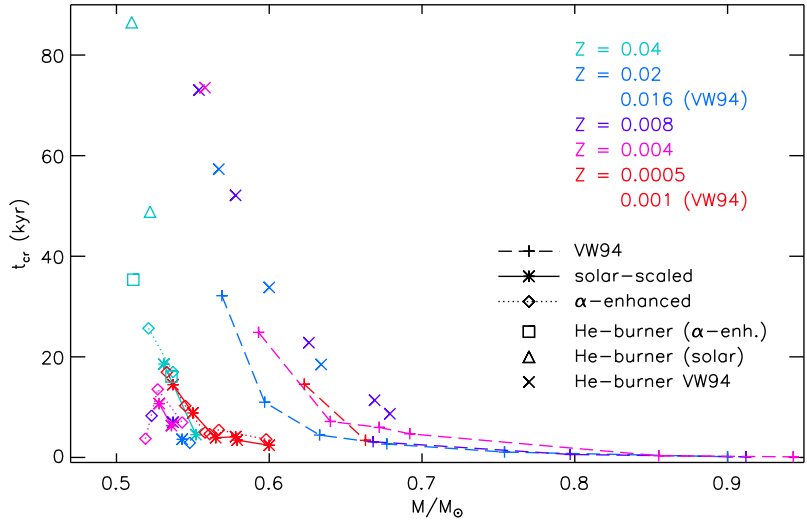}
\caption{Post-AGB crossing times (defined in the text) for our models,
  both for solar-scaled (stars) and $\alpha$-enhanced (diamonds) mixtures,
indicated by colours. For comparison, the results by \citet{vasswood:94}
are also plotted (crosses). He-burner post-AGB stars are given by the
larger open symbols, not connected by lines.}
 \end{center}
\label{f:4.12}
\end{figure}

The second timescale is the total crossing time, $t_\mathrm{cr}$,
taken from $\teff = 10^4$~K to the turn-around point at the ``knee''
of the post-AGB evolution, typically around $\teff\approx 10^5$~K. In
Fig.~\ref{f:4.12} we compare our results with those by
\citet{vasswood:94}. Our values for $t_\mathrm{tr}$ are lower
because of the faster early evolution on the post-AGB. However, some
basic features are similar: A strong increase in $t_\mathrm{cr}$ with
decreasing mass, higher values for lower metallicities at the same
mass, longer crossing times for He-burner, and a majority of values
below or around $10^4$~years. 

\begin{table}
\caption{Born-again times $t_{\mathrm{ba}}$ for models experiencing a
  (V)LTP.}
\label{t:5}
\begin{tabular}{cccccc}
\hline
$M_{\mathrm{ZAMS}}$ (M$_{\odot}$) & comp. & $M_{\mathrm{T4}}$ (M$_{\odot}$)$^a$ &
$t_{\mathrm{ba}}$ (yr) & $\log L_{\mathrm{max}}$$^b$ & $\Delta t$
(yr)$^c$\\
\hline
1.6 & III  & 0.541 & 189 & 9.386 & 10$^{-7/-2}$ \\
1.0 & V    & 0.531 & 240 & 10.05 & 10$^{-7/-2}$ \\
1.6 & V    & 0.537 & 56  & 9.959 & 10$^{-7/-2}$ \\
1.2 & VI   & 0.531 & 292 & 6.079 & 10$^{-5/-2}$ \\
1.5 & VI   & 0.537 & 302 & 6.674 & 10$^{-5/-2}$ \\
1.6 & VI   & 0.536 & 307 & 7.006 & 10$^{-5/-2}$ \\
1.5 & VII  & 0.536 & 200 & 10.34 & 10$^{-4}$ \\
1.2 & VIII & 0.538 & 254 & 6.559 & 10$^{-4/-1}$ \\
1.6 & VIII & 0.519 & 305 & 6.961 & 10$^{-6/-2}$ \\
\hline \noalign{\smallskip $^a$initial post-AGB mass; $^b$maximum
H-luminosity during the late TP; $^c$minimum time-step(s).}
\end{tabular} 
\end{table}

A summary of the various timescales is given in Tables~\ref{ta:7} and
\ref{ta:8}. The meaning of the various columns in these tables is as
follows:
  (1) initial stellar mass on ZAMS; (2)
  $\teff$ at which post-AGB evolution starts; a star indicates that this
  is not the point with a pulsation period of 100~d, but rather the first
  model after ``frozen-in'' envelope stripping; (3) stellar
  mass at $\teff = 10^4$~K; (4) transition
  time between the post-AGB start and $\teff = 10^4$~d (this is a lower
  limit for the cases marked by a star); (5) time $t_\mathrm{1}$ between
  beginning of post-AGB evolution and onset of radiation-driven wind;
  (6) time $t_\mathrm{cr}$ for crossing of HRD; (7) H- or 
  He-burner, or both, if a LTP or VLTP (indicated by column 8) happens;
  (9) end of the calculations: ``WD'' indicates that the WD cooling track
  was reached; otherwise the computations were ended due to numerical
  problems in the phase indicated.

Since some of our models experience a LTP or VLTP (see
Tables~\ref{ta:7} and \ref{ta:8}, from which also the initial post-AGB
mass $M_{\mathrm{T4}}$ is taken), we provide in Table~\ref{t:5} the
``born-again times'' $t_\mathrm{ba}$, taken as the time between the 
maximum hydrogen-luminosity after a (very) late TP has occurred
and the moment the star 
arrives again at $\log \teff = 3.8$ \citep{mba:2007}. Observationally,
\citep[e.g.][for Sakurai's object]{asplund:99},  $t_\mathrm{ba}$ is as
short as a few years, while theoretical models
\citep[e.g.][]{hbld:99,herwig:2001} predict several hundred
years. However, \citet{mbasp:2006} obtained born-again times of 5-10~years
by using an extremely fine time resolution. 
As was pointed out by \citet{herwig:2001}, care has to be taken
  with regard to the shortest time steps. The reason is that the
  mixing length theory is applicable only to  stationary convective
  situations. The adjustment of convection to changing conditions is
  not considered. Therefore, time-steps shorter than typical
  convective turn-over times are inconsistent with the assumptions of
  mixing length theory, if convective layers are changing quickly. 
  This may be the case during the fast rise of temperature due to
  violent nuclear burning, as is the case in the hot hydrogen burning
  regions of a born-again star.  The
  typical turn-over times in this phase is of order 5~minutes, or
  $10^{-5}$~years. Consequently, time-steps should be longer than this,
  and therefore \citet{mbasp:2006} set the minimum time-step allowed in
  their calculations to this value.
  For some of our models in Table~\ref{t:5}, however, we had
  to switch to even smaller time-steps to achieve numerical
  convergence. Although this happened typically only for the very
  short period around the H-luminosity maximum, and for a duration of
  order 1~hour (e.g. for the model with initially $1.6\, \msun$ and
  $M_{\mathrm{T4}}=0.558$ the time-step was $10^{-7}$~years for a total
  duration of 2.3~hrs), these models have to be taken with care. The
  total born-again time for all of the cases shown is of the order of
  several hundred years, in agreement with \citet{herwig:2001}. In
  this paper it was also demonstrated that reducing the velocity
  of convective elements, shorter born-again times can be achieved,
  while \citet{mbasp:2006} found born-again times of 5--10~years when
  going to the lower limit of acceptably small time-steps. While we in
  principle confirm this model behaviour, we had to go to even
  lower time-steps for the whole evolution past the onset of the VLTP
  and therefore, as the referee correctly  pointed out, such models
  should be considered as being unphysical.

\section{Discussion}
\label{s:discussion}

Our grid of stellar models evolving into, through, and past the Asymptotic
Giant Branch phase covers a wide range of metallicities, from
$Z=0.0005$ to $Z=0.04$ and the two standard metal distributions,
solar-scaled and $\alpha$-enhanced. The physical input -- equation of
state, opacities, nuclear reaction rates -- are completely up-to-date. Both
features are new for grids of AGB models. The masses followed the range from
$1.0$ to $6.0\,\msun$. Our grid should provide valuable input for
population studies and for synthetic AGB models, although we have not
developed new analytical functions to represent key quantities, such as
the core-mass -- luminosity relation, or the interpulse times. Previous
relations \citep{waggroen:98,itkp:2004} appear to be good first-order
approximations, which could be tuned to our models without large changes.

We specifically concentrated on a consistent treatment of carbon
enrichment of the envelope due to the third dredge-up. Carbon enrichment
is taken into account in the opacities, which have been available not only
for various $(X,Y,Z)$-mixtures, but also for changes in the C/O-ratio. To
this end new low-temperature molecular opacity tables were calculated and
included in the stellar evolution program. The increase in carbon leads to
lower effective temperature, mainly by the increase in total
metallicity. We found that at same metallicity, an increase in C/O can
initially lead to lower opacities and thus higher effective temperatures as
compared to a solar-scaled mixture. At C/O~$\gtrsim 2$ finally higher
opacities are reached. 

For the crucial mass loss of highly evolved AGB stars we used an empirical
mass loss formula for oxygen stars \citep{vlczl:2005} and a theoretical
one for carbon stars \citep{wswas:2002}. They are both similar in the
order of magnitude of the mass loss and used only if the pulsation period is
above 400~d. The switching-on of a strong mass loss from an (enhanced)
Reimers wind happens quite suddenly at a pulsation period of 400~d, with a
definite decrease in effective 
temperature. This is to be considered as a consequence of our specific
mass loss description. Carbon-enhancement together with increasing
opacities and decreasing effective temperature lead to a strong mass loss,
which eventually terminates in a strong ``superwind''.

Third dredge-up is obtained by assuming overshooting (implemented by a
diffusion approach), the extent of which is set by the free
parameter $f$ of this prescription (Eq.~\ref{e:ov}). 
We used a numerical value of $0.016$ obtained from other,
independent calibrations (open clusters, upper main-sequence). 
Although a very similar value has also been used in earlier AGB models
\citep{hbld:99,herwig:2000}, specific comparison with the results of
nuclear processing indicate that it may vary within AGB stars: being
possibly smaller in the pulse-driven convection zone \citep{lhlgs:2003}, and
larger at the bottom of the convective envelope
\citep{hll:2003}. In spite of these hints we decided to refrain from 
varying the overshooting parameter.
Nevertheless, the result of the third dredge-up, obtained with
this prescription and lifetimes for carbon-stars are in good agreement with
observations. Such agreement has been reached before only with tunable
synthetic models.

On the other hand, our predicted initial-final mass relation, though
overall in very good agreement with a recent empirical determination,
shows that the hydrogen-free core grows less than observed for the lowest
initial stellar masses. This is due to our application of overshooting to
all convective boundaries. A reduction of overshooting at the base of the
pulse-driven convection zone seems to be supported. 

A severe problem -- not specific to our calculations -- are the
convergence failures for highly evolved AGB 
stars. In particular for the higher masses and lower metallicities they
partially prevent useful calculations. They are physically connected to a
dominance of radiation pressure in the lower convective envelope
\citep{wf:86}, which also leads to non-physical, super-sonic convective
velocities \citep{wag:96} within the mixing-length theory. These
convergence problems appear in many 
modern AGB calculations, such as \citet{karakas:2003}, which has been the
most extensive and modern grid so far. A proper treatment, guided by
hydrodynamical calculations, is urgently asked for. Since we did not use
any of the ``recipes'' to somehow bypass this critical phase, the number
of reasonably complete AGB models is $\approx 100$, out of the 110 models our grid
in total comprises. For the post-AGB phase, the number even reduces to 60,
out of which only 30 could be followed from the ZAMS to the WD stage
without interruption. The remaining 30 cases were obtained by artificially
stripping off the residual envelope in the early AGB-transition
phase. All these cases correspond to an initial mass of up to $2\,\msun$ and
a post-AGB mass below $0.60\,\msun$. These models constitute a useful
extension of previous post-AGB models by \citet{vasswood:94} and
\citet{bloecker:95b} to lower post-AGB masses.

Since we aimed at treating the effect of carbon-enrichment of the envelope
as consistently as possible, it is interesting to compare the effective
temperatures of our models with those obtained with the synthetic models
by \citet{magi:2007}, who had a similar fully consistent treatment of AGB
evolution in mind, and calibrated their synthetic models to observed
carbon-star luminosity functions. We specifically looked into their model of
$1.8\,\msun$ and $Z=0.008$ (shown in Fig.~5 of that paper). In fact, the
$\teff$ evolution during the AGB-phase is very similar: Both models start
at $\log\teff \approx 3.6$ in the early interpulse phases and then become
cooler down to $\log\teff \approx 3.4$ and below during the last
TP. However, our model shows a more gradual decrease because of the better
resolution of C/O-variations in the WSU molecular opacities. There are, of
course, differences, as well. Our model experiences only 9 TPs (plus 1
estimated final TP), while theirs has 32, with C/O~$>1$ reached at
TP~19. The final C/O-ratio is close to 3. Our model ends at C/O$=2.004$,
and turns into a C-star after TP~4. This is due to the different dredge-up
efficiencies and mass loss descriptions. A similar agreement is found for
the $M=4\,\msun$ model, 
except that our model does not experience HBB and therefore the increase
in $\teff$ due to a decrease in C/O is not taking place. Before that
event, during the initial 3du, both models also show a similar
$\teff$-development. \citet{magi:2007} emphasize that their improved
treatment of molecular opacities leads to much better agreement with
observed colors of AGB stars \citep{mgbgsg:2008}.
Together with the fact that our full AGB models are
the first to reproduce the carbon-star lifetimes as derived by
\citet{gima:2007}, we conclude that overall our full models agree well
with these calibrated and successful synthetic models, although they were
computed with pre-defined physical input. 

In the present work, a detailed investigation of nucleosynthesis and
chemical yields was ignored \citep[see][for such
  details]{karakas:2003}. However, we can confirm that our models display
the occurrence of $^{13}\mathrm{C}$-pockets as a consequence of 3du and
convective mixing. Therefore, the necessary precondition for s-process
nucleosynthesis is given, but a detailed analysis with a post-processing
network is still needed. 

We have used the most up-to-date and self-consistent physical approach to
full AGB models. To our surprise, our models often agree better with the
older \citet{vasswood:93} models than with the newer ones by
Karakas. These two sets were actually calculated with the same code, but
in different versions, separated by a decade of development. We think that
this indicates that the numerical and technical aspects of implementing
the relevant physics for AGB-stars is still an important factor, such that
future work in this direction is as important as further improvements in
the physics itself.


\begin{acknowledgements}
We have presented in this paper mainly the results of the PhD thesis
by Agis~Kitsikis, to whom we are obliged for allowing us to use
his calculations. We thank P.~Marigo for making available her
molecular opacities to us, and M.~Miller~Bertolami for model
comparisons, additional calculations, and helpful discussions. The
(anonymous) referee provided an exceptionally expert and helpful
review which corrected several misunderstandings on our side, and for
which we are very grateful.
\end{acknowledgements}
\clearpage

\bibliographystyle{aa}
\bibliography{agb}

\clearpage

\begin{appendix}

\section{Evolution up to the first TP}
\label{a:time}

\begin{table*}[htp!]
\caption{Solar-scaled metallicity models
  from the ZAMS until the 1$^{\mathrm{st}}$ TP. }
\label{ta:1}
\begin{tabular}{ccccccccccccccc}
\multicolumn{6}{c}{$Z = 0.02$ / solar} & 
&  & \multicolumn{7}{c}{$Z = 0.008$ / solar} \\
\hline
$M_{\mathrm{ZAMS}}$ & $t_{\mathrm{MS}}$$^a$ & $t_{\mathrm{RGB}}$
& $t_{\mathrm{He-b}}$ & $t_{\mathrm{EAGB}}$ & $M_{\mathrm{tot}}$(1)$^b$ & $M_\mathrm{c}$(1)$^c$ & 
& $M_{\mathrm{ZAMS}}$ & $t_{\mathrm{MS}}$ & $t_{\mathrm{RGB}}$
& $t_{\mathrm{He-b}}$ & $t_{\mathrm{EAGB}}$ & $M_{\mathrm{tot}}$(1) & $M_\mathrm{c}$(1) \\
\hline
1.0 & 8616.2 & 3350.1 & 110.17 & 12.971 & 0.709 & 0.501 &  & 1.0 & 6751.9 & 2521.0 & 104.48 & 10.369 & 0.742 & 0.513 \\
1.2 & 4473.9 & 1555.6 & 106.47 & 12.368 & 0.981 & 0.509 &  & 1.2 & 3600.5 & 1182.0 & 97.963 & 10.474 & 0.999 & 0.521 \\
1.5 & 2560.8 & 242.28 & 108.73 & 11.645 & 1.353 & 0.511 &  & 1.5 & 2097.9 & 193.39 & 99.367 & 9.8133 & 1.366 & 0.521 \\
1.6 & 2091.5 & 172.58 & 110.80 & 12.243 & 1.476 & 0.508 &  & 1.6 & 1725.1 & 137.98 & 100.87 & 11.363 & 1.490 & 0.516 \\
1.8 & 1465.8 & 99.847 & 130.85 & 13.305 & 1.595 & 0.496 &  & 1.8 & 1223.8 & 78.851 & 128.10 & 11.802 & 1.637 & 0.497 \\
2.0 & 1075.0 & 58.364 & 177.34 & 15.417 & 1.895 & 0.478 &  & 2.0 & 900.54 & 32.110 & 236.41 & 14.571 & 1.941 & 0.464 \\
2.6 & 510.25 & 12.302 & 140.08 & 10.175 & 2.530 & 0.518 &  & 2.6 & 448.03 & 9.9558 & 106.62 & 6.7430 & 2.506 & 0.569 \\
3.0 & 344.35 & 7.1561 & 83.500 & 5.7529 & 2.900 & 0.596 &  & 3.0 & 309.24 & 5.9283 & 60.199 & 3.7127 & 2.863 & 0.672 \\
4.0 & 161.46 & 2.7018 & 28.711 & 1.8755 & 3.821 & 0.765 &  & 4.0 & 152.27 & 2.3324 & 23.405 & 1.2466 & 3.812 & 0.801 \\
5.0 & 92.624 & 1.3843 & 13.577 & 0.9122 & 4.800 & 0.830 &  & 5.0 & 90.826 & 1.2398 & 12.184 & 0.6929 & 4.776 & 0.868 \\
6.0 & 60.471 & 0.8147 & 8.1598 & 0.4890 & 5.767 & 0.928 &  & 6.0 & 60.976 & 0.7491 & 7.7504 & 0.4100 & 5.711 & 0.991 \\
\noalign{\smallskip} \hline
\noalign{\medskip}
\multicolumn{6}{c}{$Z = 0.04$ / solar} & 
&  & \multicolumn{7}{c}{$Z = 0.004$ / solar} \\
\hline
$M_{\mathrm{ZAMS}}$ & $t_{\mathrm{MS}}$ & $t_{\mathrm{RGB}}$
& $t_{\mathrm{He-b}}$ & $t_{\mathrm{EAGB}}$ & $M_{\mathrm{tot}}$(1) & $M_\mathrm{c}$(1) & 
& $M_{\mathrm{ZAMS}}$ & $t_{\mathrm{MS}}$ & $t_{\mathrm{RGB}}$
& $t_{\mathrm{He-b}}$ & $t_{\mathrm{EAGB}}$ & $M_{\mathrm{tot}}$(1) & $M_\mathrm{c}$(1) \\
\hline
1.0 & 8648.4 & 3974.3 & 119.31 & 14.213 & 0.682 & 0.504 &  & 1.0 & 5771.4 & 2050.5 & 92.852 & 10.965 & 0.776 & 0.517 \\
1.2 & 4788.6 & 1527.4 & 119.81 & 12.305 & 0.963 & 0.513 &  & 1.2 & 3141.2 & 966.74 & 94.451 & 9.3231 & 1.026 & 0.525 \\
1.5 & 2697.1 & 231.58 & 122.80 & 12.425 & 1.349 & 0.513 &  & 1.5 & 1847.2 & 164.02 & 94.336 & 9.2462 & 1.383 & 0.527 \\
1.6 & 2196.0 & 167.42 & 124.74 & 14.042 & 1.476 & 0.510 &  & 1.6 & 1525.1 & 117.01 & 103.22 & 9.0807 & 1.505 & 0.521 \\
1.8 & 1527.7 & 96.664 & 158.38 & 13.966 & 1.611 & 0.496 &  & 1.8 & 1088.4 & 65.247 & 131.73 & 11.283 & 1.672 & 0.499 \\
2.0 & 1108.6 & 54.286 & 226.81 & 17.353 & 1.903 & 0.481 &  & 2.0 & 812.67 & 27.464 & 199.22 & 11.361 & 1.935 & 0.488 \\
2.6 & 509.95 & 13.146 & 140.91 & 10.013 & 2.518 & 0.528 &  & 2.6 & 410.03 & 8.5853 & 82.059 & 4.0737 & 2.472 & 0.639 \\
3.0 & 337.90 & 7.6362 & 83.253 & 6.2347 & 2.893 & 0.596 &  & 3.0 & 286.74 & 5.2255 & 50.584 & 2.3417 & 2.834 & 0.744 \\
4.0 & 151.95 & 2.8971 & 29.886 & 2.1740 & 3.826 & 0.750 &  & 4.0 & 145.00 & 2.1178 & 20.483 & 1.1477 & 3.801 & 0.809 \\
5.0 & 84.428 & 1.4122 & 13.929 & 1.0935 & 4.791 & 0.810 &  & 5.0 & 88.228 & 1.1468 & 11.488 & 0.5852 & 4.775 & 0.894 \\
6.0 & 53.949 & 0.8091 & 7.9334 & 0.5929 & 5.651 & 0.893 &  & 6.0 &
60.070 & 0.7068 & 7.5367 & 0.3372 & 5.700 & 1.060 \\
\noalign{\smallskip} \hline
\noalign{\medskip}
\end{tabular}
\end{table*}

\begin{table*}[h]
\begin{tabular}{ccccccc}
\multicolumn{6}{c}{$Z = 0.0005$ / solar} \\
\hline
$M_{\mathrm{ZAMS}}$ & $t_{\mathrm{MS}}$ & $t_{\mathrm{RGB}}$
& $t_{\mathrm{He-b}}$ & $t_{\mathrm{EAGB}}$ & $M_{\mathrm{tot}}$(1) & $M_\mathrm{c}$(1) \\
\hline
1.0 & 5041.2 & 1262.2 & 83.219 & 9.4308 & 0.847 & 0.523  \\
1.2 & 2662.3 & 727.55 & 86.164 & 7.4351 & 1.078 & 0.537  \\
1.5 & 1560.7 & 158.70 & 73.039 & 6.7829 & 1.351 & 0.570  \\
1.6 & 1278.6 & 104.02 & 87.687 & 6.7574 & 1.521 & 0.561  \\
1.8 & 912.34 & 46.277 & 147.74 & 6.5325 & 1.701 & 0.551  \\
2.0 & 682.51 & 22.101 & 127.21 & 5.6236 & 1.906 & 0.584  \\
2.6 & 351.50 & 7.1310 & 59.160 & 2.4622 & 2.464 & 0.732  \\
3.0 & 250.82 & 4.4426 & 39.685 & 1.6650 & 2.857 & 0.780  \\
4.0 & 132.50 & 1.8581 & 17.728 & 0.8079 & 3.867 & 0.848  \\
5.0 & 83.085 & 1.0331 & 9.9502 & 0.4567 & 4.875 & 0.937  \\
6.0 & 57.746 & 0.6695 & 6.3498 & 0.4317 & 4.208 & 1.128  \\
\noalign{\smallskip} \hline
\noalign{\medskip
$^a$ All times are given in 10$^6$ yrs; 
$^b$ total mass at 1$^{\mathrm{st}}$ TP in solar unit;
$^c$ core mass at 1$^{\mathrm{st}}$ TP.}
\end{tabular}
\end{table*}

\begin{table*}[h]
\caption{Same as for Table~\ref{ta:1} but for the $\alpha$-enhanced
  metallicity models.}
\label{ta:2}
\begin{tabular}{ccccccccccccccc}
\multicolumn{6}{c}{$Z = 0.02$ / $\alpha$-enhanced} & 
&  & \multicolumn{7}{c}{$Z = 0.008$ / $\alpha$-enhanced} \\
\hline
$M_{\mathrm{ZAMS}}$ & $t_{\mathrm{MS}}$ & $t_{\mathrm{RGB}}$
& $t_{\mathrm{He-b}}$ & $t_{\mathrm{EAGB}}$ & $M_{\mathrm{tot}}$(1) & $M_\mathrm{c}$(1) & 
& $M_{\mathrm{ZAMS}}$ & $t_{\mathrm{MS}}$ & $t_{\mathrm{RGB}}$
& $t_{\mathrm{He-b}}$ & $t_{\mathrm{EAGB}}$ & $M_{\mathrm{tot}}$(1) & $M_\mathrm{c}$(1) \\
\hline
1.0 & 8442.0 & 3065.5 & 112.02 & 10.614 & 0.713 & 0.502 & & 1.0 & 6633.4 & 2419.4 & 98.966 & 10.164 & 0.755 & 0.508 \\
1.2 & 4153.6 & 1607.1 & 105.88 & 10.772 & 0.985 & 0.508 & & 1.2 & 3441.4 & 1194.1 & 98.397 & 9.2639 & 1.011 & 0.517 \\
1.5 & 2287.8 & 304.00 & 99.117 & 11.415 & 1.346 & 0.514 & & 1.5 & 1971.4 & 218.60 & 97.908 & 8.8516 & 1.360 & 0.526 \\
1.6 & 1865.5 & 212.48 & 109.24 & 9.9499 & 1.468 & 0.513 & & 1.6 & 1619.9 & 154.04 & 100.92 & 9.3640 & 1.483 & 0.522 \\
1.8 & 1465.8 & 99.847 & 130.85 & 13.305 & 1.569 & 0.501 & & 1.8 & 1147.0 & 86.083 & 118.83 & 11.511 & 1.616 & 0.505 \\
2.0 & 1075.0 & 58.364 & 177.34 & 15.423 & 1.867 & 0.488 & & 2.0 & 850.65 & 35.418 & 238.79 & 15.050 & 1.939 & 0.464 \\
2.6 & 510.25 & 12.297 & 140.08 & 10.175 & 2.529 & 0.513 & & 2.6 & 421.73 & 10.360 & 111.36 & 5.6708 & 2.503 & 0.578 \\
3.0 & 344.35 & 7.1561 & 83.500 & 5.7529 & 2.909 & 0.585 & & 3.0 & 292.00 & 6.0452 & 62.486 & 3.2070 & 2.857 & 0.684 \\
4.0 & 146.84 & 2.7657 & 28.801 & 1.9129 & 3.823 & 0.766 & & 4.0 & 144.93 & 2.3134 & 22.858 & 1.3036 & 3.811 & 0.799 \\
5.0 & 85.371 & 1.3850 & 14.328 & 0.8164 & 4.802 & 0.846 & & 5.0 & 87.124 & 1.2145 & 121.36 & 0.6650 & 4.771 & 0.875 \\
6.0 & 56.370 & 0.7959 & 8.3754 & 0.4509 & 5.715 & 0.946 & & 6.0 & 58.838 & 0.7307 & 7.5395 & 0.4267 & 5.713 & 0.996 \\
\noalign{\smallskip} \hline
\noalign{\medskip}
\multicolumn{6}{c}{$Z = 0.04$ / $\alpha$-enhanced} & 
&  & \multicolumn{7}{c}{$Z = 0.004$ / $\alpha$-enhanced} \\
\hline
$M_{\mathrm{ZAMS}}$ & $t_{\mathrm{MS}}$ & $t_{\mathrm{RGB}}$
& $t_{\mathrm{He-b}}$ & $t_{\mathrm{EAGB}}$ & $M_{\mathrm{tot}}$(1) & $M_\mathrm{c}$(1) & 
& $M_{\mathrm{ZAMS}}$ & $t_{\mathrm{MS}}$ & $t_{\mathrm{RGB}}$
& $t_{\mathrm{He-b}}$ & $t_{\mathrm{EAGB}}$ & $M_{\mathrm{tot}}$(1) & $M_\mathrm{c}$(1) \\
\hline
1.0 & 8445.6 & 3482.5 & 111.90 & 13.314 & 0.678 & 0.505 & & 1.0 & 5696.1 & 2004.1 & 94.355 & 9.5685 & 0.784 & 0.515 \\
1.2 & 4181.9 & 1728.5 & 107.84 & 12.716 & 0.964 & 0.512 & & 1.2 & 3053.4 & 974.38 & 93.588 & 8.6953 & 1.032 & 0.523 \\
1.5 & 2293.0 & 315.64 & 108.53 & 11.806 & 1.333 & 0.518 & & 1.5 & 1781.6 & 173.96 & 93.097 & 8.6722 & 1.379 & 0.531 \\
1.6 & 1861.1 & 222.36 & 109.39 & 12.835 & 1.462 & 0.514 & & 1.6 & 1468.9 & 124.17 & 99.306 & 8.6461 & 1.498 & 0.529 \\
1.8 & 1291.9 & 125.72 & 131.82 & 12.497 & 1.562 & 0.504 & & 1.8 & 1047.6 & 67.684 & 126.58 & 10.905 & 1.662 & 0.506 \\
2.0 & 1108.6 & 54.286 & 226.81 & 17.853 & 1.855 & 0.494 & & 2.0 & 782.24 & 28.666 & 197.48 & 10.426 & 1.933 & 0.494 \\
2.6 & 509.95 & 13.146 & 140.91 & 10.013 & 2.520 & 0.515 & & 2.6 & 392.18 & 8.6687 & 79.021 & 3.9082 & 2.469 & 0.648 \\
3.0 & 337.90 & 7.6362 & 83.252 & 6.2347 & 2.908 & 0.574 & & 3.0 & 277.07 & 5.2243 & 48.452 & 2.5145 & 2.833 & 0.744 \\
4.0 & 132.12 & 3.0023 & 32.044 & 2.3654 & 3.804 & 0.749 & & 4.0 & 140.83 & 2.0926 & 21.321 & 0.9967 & 3.801 & 0.827 \\
5.0 & 74.606 & 1.4437 & 15.476 & 0.9332 & 4.753 & 0.830 & & 5.0 & 86.130 & 1.1276 & 11.536 & 0.5443 & 4.786 & 0.908 \\
6.0 & 48.420 & 0.8106 & 8.8116 & 0.4817 & 5.397 & 0.935 & & 6.0 & 58.868 & 0.6937 & 7.1015 & 0.3700 & 5.740 & 1.050 \\
\noalign{\smallskip} \hline
\noalign{\medskip}
\end{tabular}
\end{table*}

\begin{table*}[htp!]
\begin{tabular}{ccccccc}
\multicolumn{7}{c}{$Z = 0.0005$ / $\alpha$-enhanced} \\
\hline
$M_{\mathrm{ZAMS}}$ & $t_{\mathrm{MS}}$ & $t_{\mathrm{RGB}}$
& $t_{\mathrm{He-b}}$ & $t_{\mathrm{EAGB}}$ & $M_{\mathrm{tot}}$(1) & $M_\mathrm{c}$(1) \\
\hline
 1.0 & 5034.3 & 1246.6 & 82.220 & 9.1452 & 0.843 & 0.527 \\
 1.2 & 2650.9 & 725.59 & 80.412 & 8.3182 & 1.077 & 0.538 \\
 1.5 & 1536.4 & 142.67 & 87.219 & 6.1783 & 1.409 & 0.559 \\
 1.6 & 1268.5 & 104.90 & 91.057 & 5.9047 & 1.523 & 0.562 \\
 1.8 & 904.79 & 47.096 & 144.84 & 6.4154 & 1.700 & 0.554 \\
 2.0 & 676.61 & 22.188 & 126.53 & 5.4208 & 1.903 & 0.588 \\
 2.6 & 348.59 & 7.1208 & 59.499 & 2.3104 & 2.467 & 0.737 \\
 3.0 & 249.01 & 4.4098 & 38.353 & 1.7691 & 2.859 & 0.777 \\
 4.0 & 131.74 & 1.8480 & 17.972 & 0.7419 & 3.871 & 0.862 \\
 5.0 & 82.733 & 1.0312 & 9.8540 & 0.4491 & 4.878 & 0.943 \\
 6.0 & 57.554 & 0.6858 & 6.2844 & - & - \\
\end{tabular}
\end{table*}

\begin{figure*}[h]
\includegraphics[angle=90,scale=0.3]{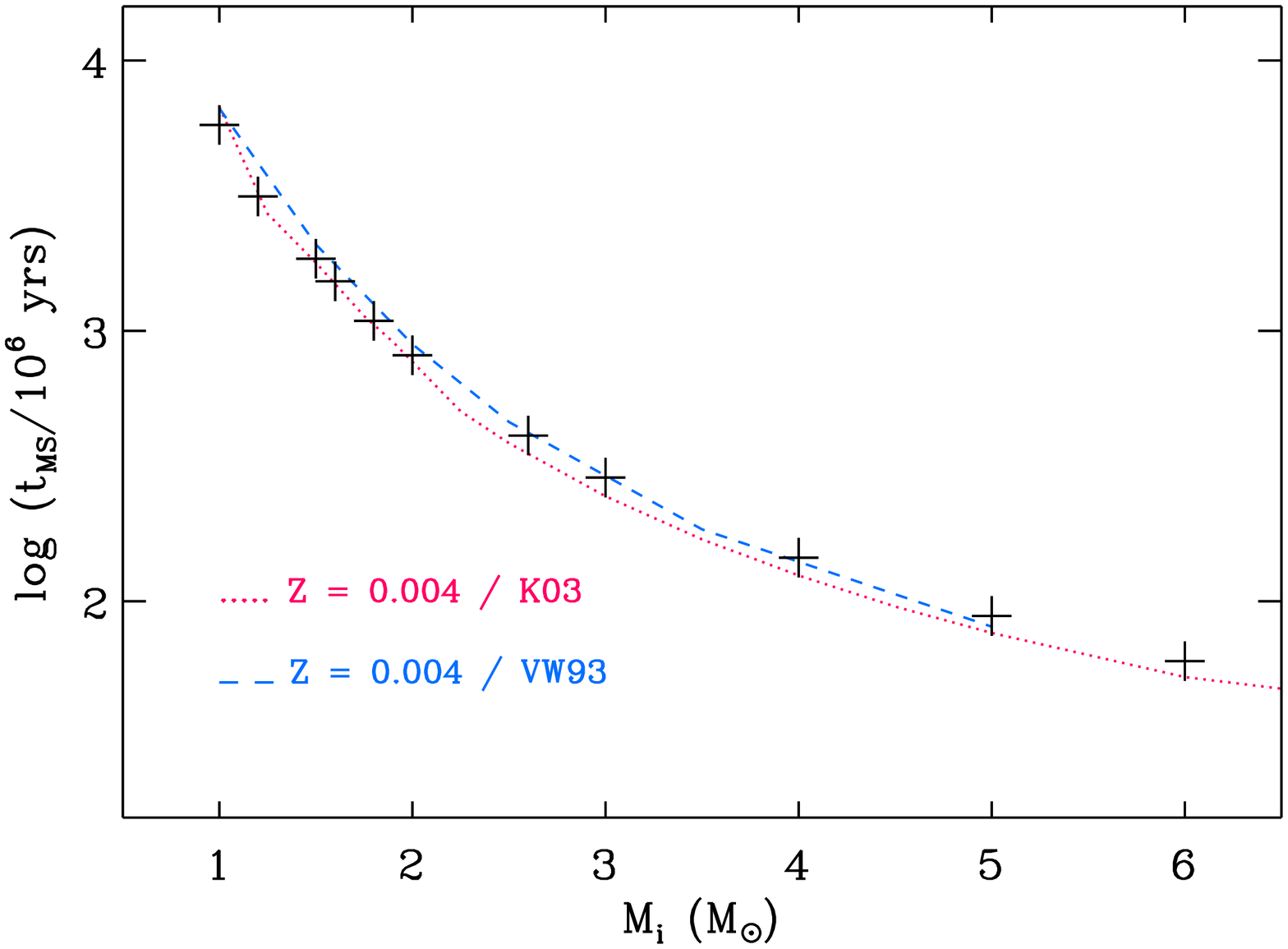}
\includegraphics[angle=90,scale=0.3]{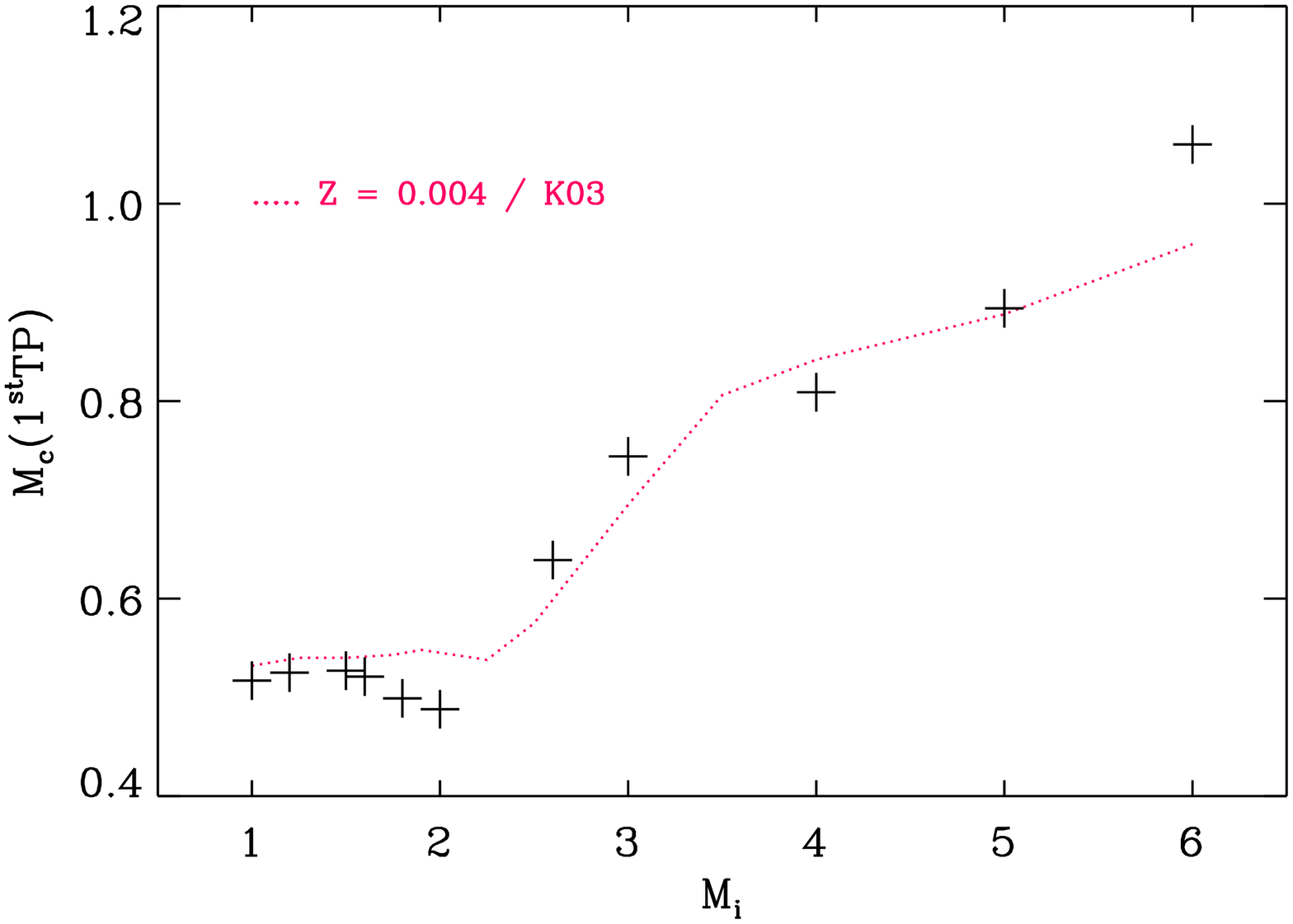} \\
\includegraphics[angle=90,scale=0.3]{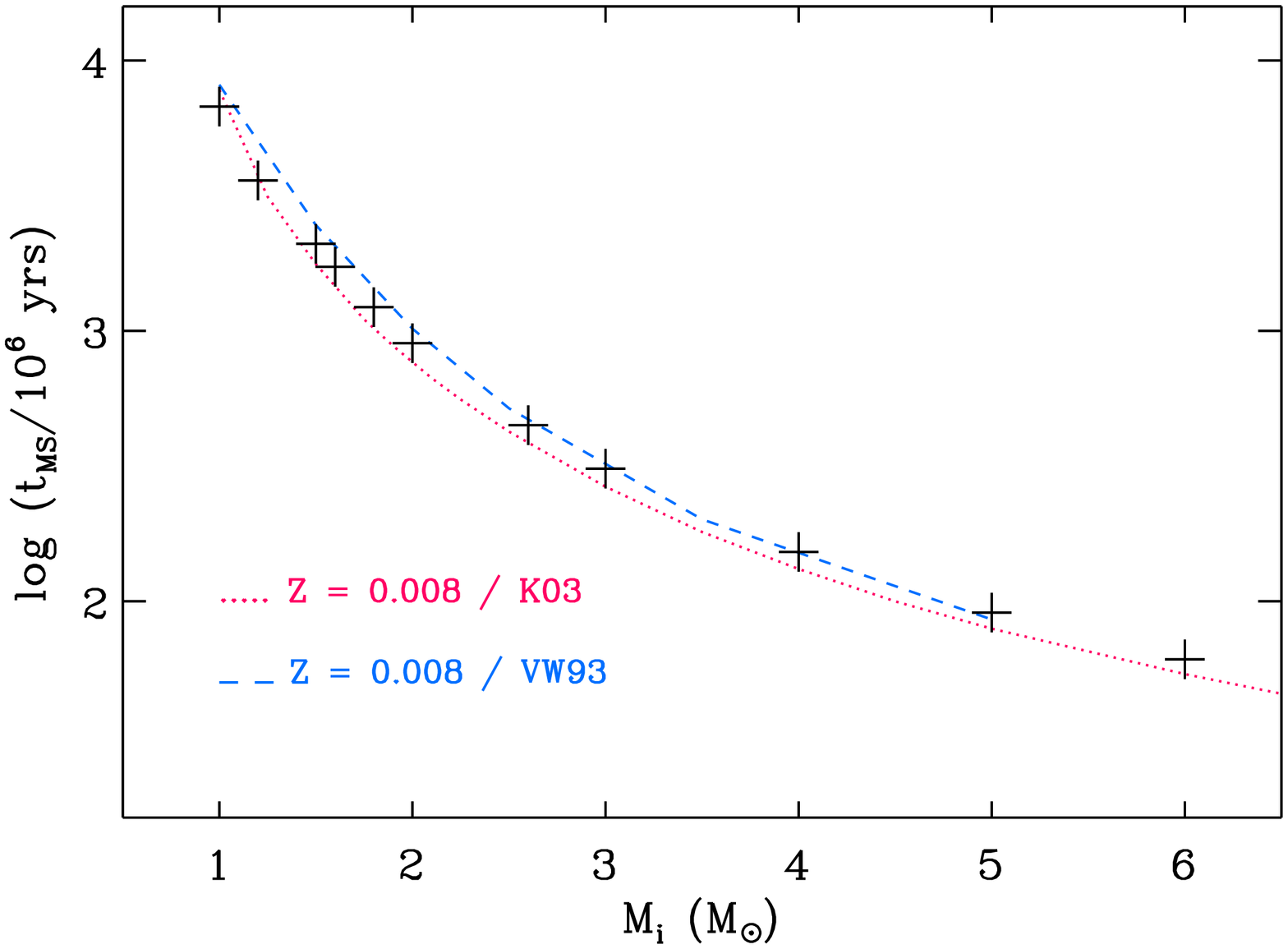}
\includegraphics[angle=90,scale=0.3]{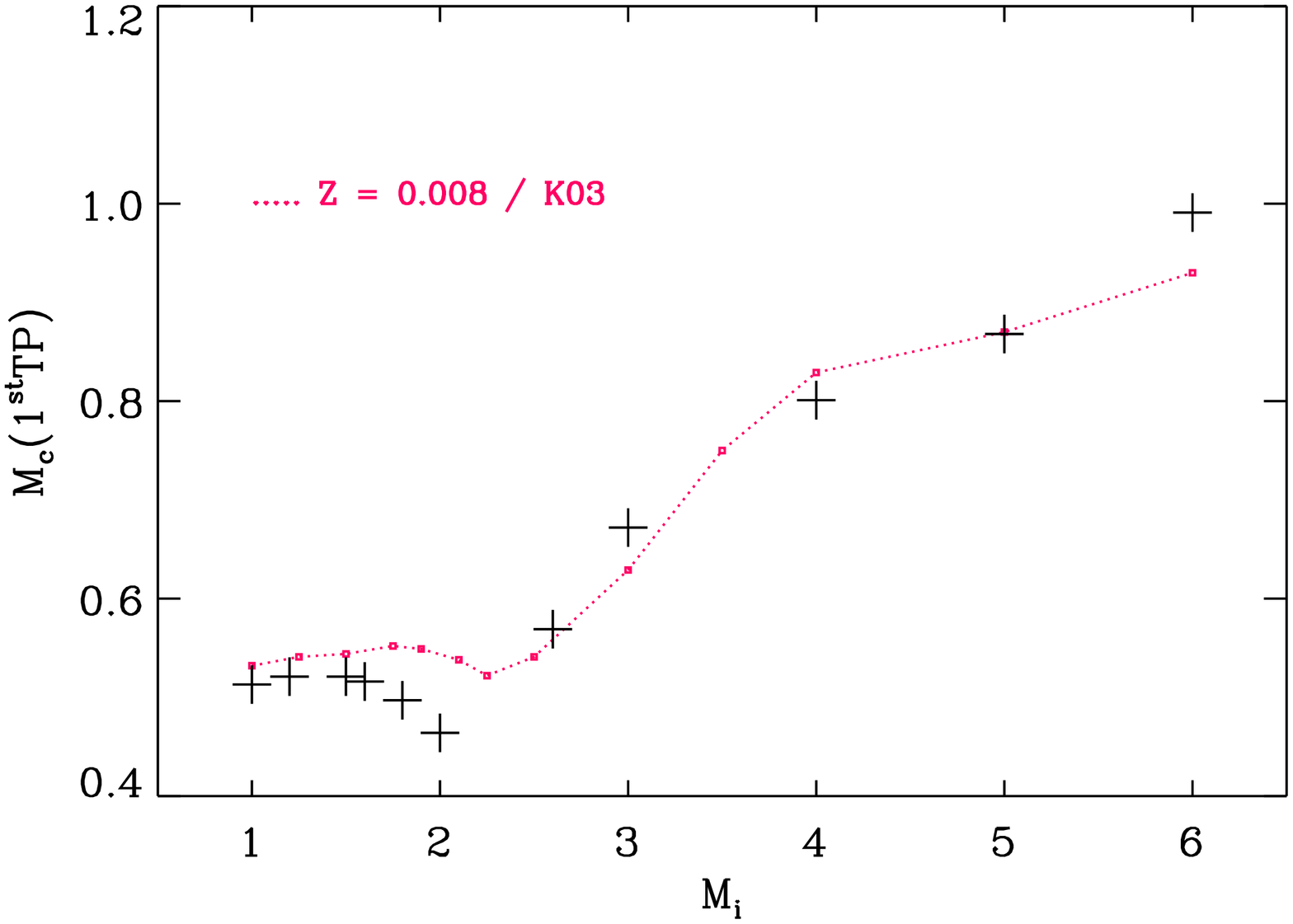} \\
\includegraphics[angle=90,scale=0.3]{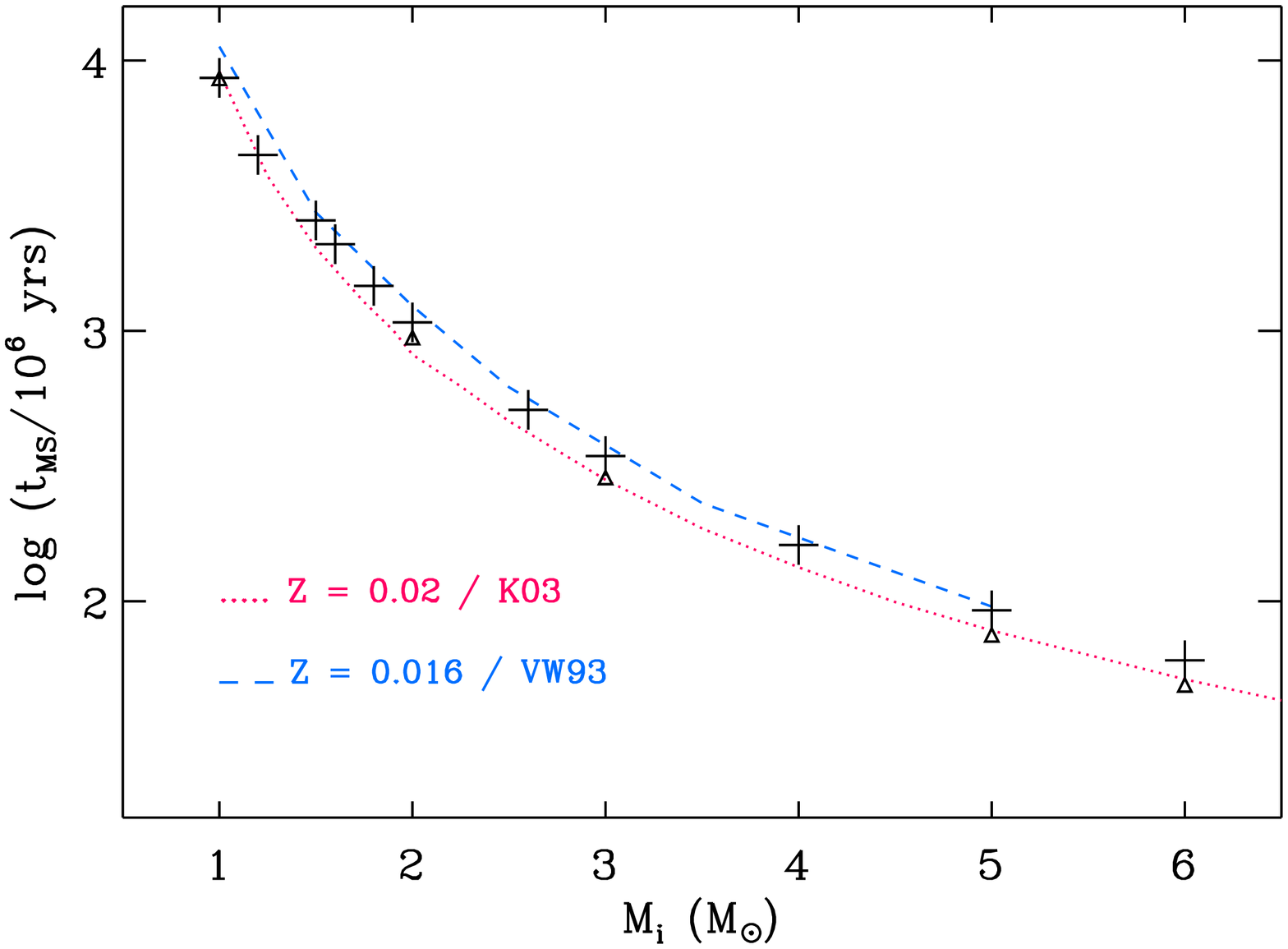}
\includegraphics[angle=90,scale=0.3]{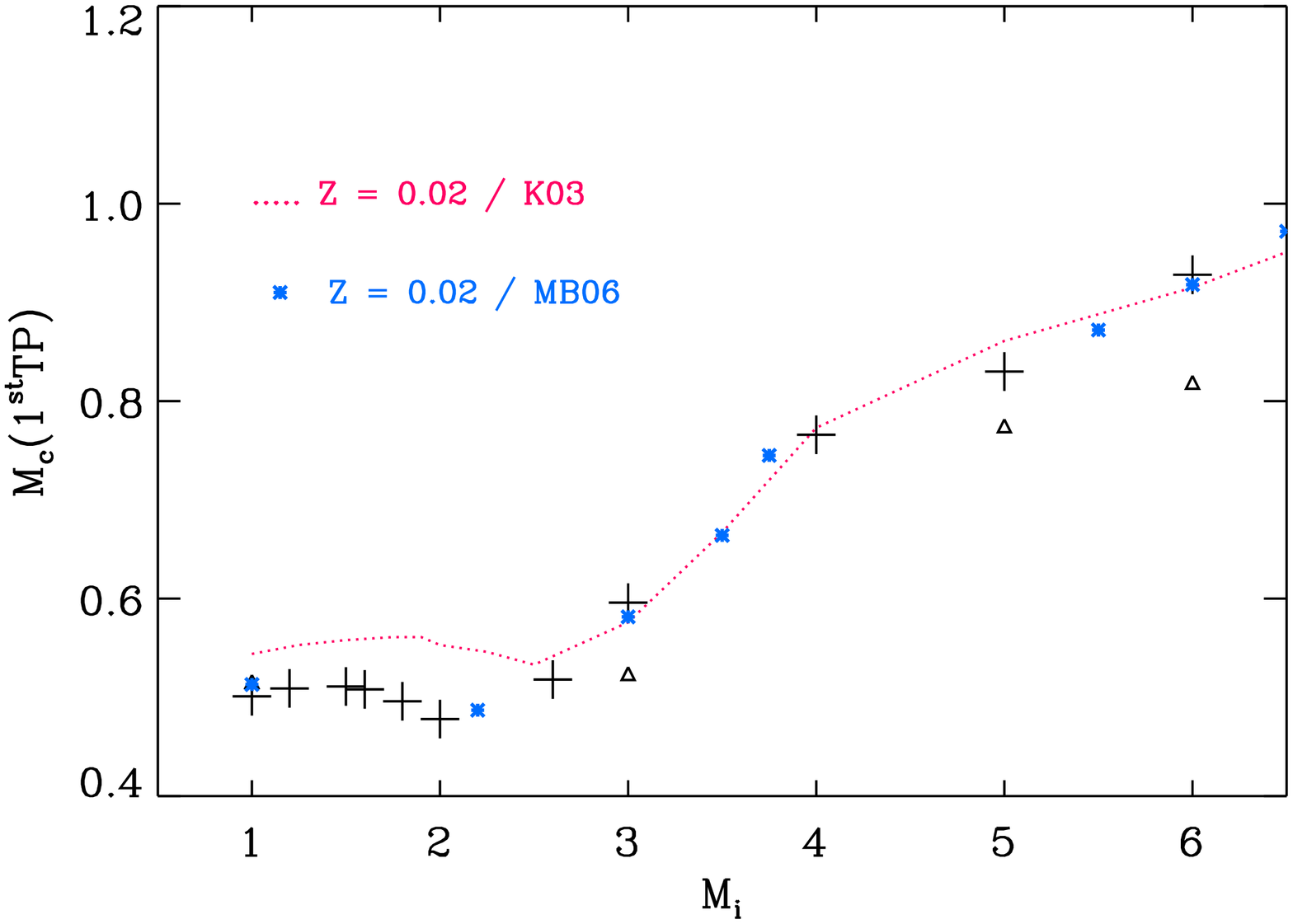}
\caption{Main sequence lifetimes (left) and core mass at the first
  thermal pulse (right) for three chemical compositions with
  $Z=0.004$, 0.008, 0.02 (from top to bottom) of our models (crosses)
  in comparison with literature values from \citet[K03]{karakas:2003},
\citet[VW93]{vasswood:93}, and Miller-Bertolami (private
communication, 2008, MB08). }
\label{fa:1}
\end{figure*}

\clearpage

\section{Summary of AGB properties}
\label{a:summary}

\begin{table*}[h]
\caption{Same as Table~\ref{t:4}, but for mixtures I and II
  (Table~\ref{t:1})   with $Z=0.04$.}
\label{ta:3} 
\begin{tabular}{ccccccccccrrr}
\hline
\noalign{\smallskip}
$M_{\mathrm{ZAMS}}$ & No & $M_\mathrm{c}$(1) & $M_\mathrm{c}$(3du)
& No & C/O$_{\mathrm{f}}$ & $M$(f) & $M_\mathrm{c}$(f)
& No & No & \multicolumn{3}{c}{TP-AGB lifetime (10$^{3}$ yrs)} \\
\cline{11-13}
& TPs & & & TP$_{\mathrm{3du}}$(i) & & & & TP$_{\mathrm{HBB}}$(i) &
TP$_{\mathrm{HBB}}$(f) & $t_{\mathrm{mod}}$ & $\approx t_{\mathrm{e}}$
(TPs) & $t_{\mathrm{TP}}$  \\
\noalign{\smallskip}
\hline
\noalign{\smallskip}
\multicolumn{13}{c}{$Z = 0.04$ / solar (mixture I)} \\
\noalign{\smallskip}
\hline
\noalign{\smallskip}
1.0 & 2 & 0.504 & - & - & 0.381 & 0.514 & 0.510 & - & - & {244.754} & {-} & {245} \\
1.2 & 4 & 0.513 & - & - & 0.356 & 0.524 & 0.521 & - & - & {311.523} & {-} & {311} \\
1.5 & 6 & 0.513 & - & - & 0.331 & 0.533 & 0.531 & - & - & {523.002} & {-} & {523} \\
1.6 & 7 & 0.510 & - & - & 0.334 & 0.544 & 0.536 & - & - & {705.362} & {0.5 (0)} & {706} \\
1.8 & 10 & 0.496 & - & - & 0.323 & 0.539 & 0.537 & - & - & {1107.536} & {-} & {1108} \\
2.0 & 13 & 0.481 & - & - & 0.322 & 0.552 & 0.552 & - & - & {1876.474} & {-} & {1876} \\
2.6 & 15 & 0.528 & 0.566 & 10 & 0.449 & 1.037 & 0.587 & - & - & {999.657} & {28 (0)} & {1 028} \\
3.0 & 10 & 0.596 & 0.605 & 4 & 0.507 & 0.976 & 0.628 & - & - & {402.750} & {14 (0)} & {417} \\
4.0 & 8 & 0.750 & 0.751 & 2 & 0.583 & 1.988 & 0.751 & - & - & {86.755} & {31 (2)} & {118} \\
5.0 & 16 & 0.810 & 0.812 & 2 & 0.444 & 2.573 & 0.829 & - & - & {94.703} & {44 (5)} & {139} \\
6.0 & 26 & 0.893 & 0.893 & 1 & 0.455 & 3.333 & 0.908 & 6 & 17 & {74.350} & {48 (12)} & {86} \\
\noalign{\smallskip}
\hline\noalign{\smallskip}
\multicolumn{13}{c}{Z$ = 0.04$ / $\alpha$-enhanced (mixture II)} \\
\noalign{\smallskip}
\hline
\noalign{\smallskip}
1.0 & 2 & 0.505 & - & - & 0.163 & 0.516 & 0.510 & - & - & {235.332} & {-} & {235} \\
1.2 & 3 & 0.512 & - & - & 0.154 & 0.523 & 0.520 & - & - & {333.567} & {-} & {334} \\
1.5 & 5 & 0.518 & - & - & 0.139 & 0.539 & 0.535 & - & - & {466.447} & {-} & {466} \\
1.6 & 6 & 0.514 & - & - & 0.150 & 0.698 & 0.536 & - & - & {576.262} & {16 (0)} & {592} \\
1.8 & 8 & 0.504 & - & - & 0.134 & 0.538 & 0.536 & - & - & {856.256} & {-} & {856} \\
2.0 & 13 & 0.494 & - & - & 0.135 & 0.590 & 0.552 & - & - & {1 394.989} & {4 (0)} & {1399} \\
2.6 & 16 & 0.515 & 0.558 & 10 & 0.432 & 0.745 & 0.580 & - & - & {1 288.965} & {5 (0)} & {1294} \\
3.0 & 13 & 0.574 & 0.587 & 5 & 0.442 & 0.912 & 0.613 & - & - & {617.336} & {9 (0)} & {626} \\
4.0 & 8 & 0.750 & 0.749 & 1 & 0.234 & 1.795 & 0.763 & - & - & {96.381} & {26 (1)} & {122} \\
5.0 & 15 & 0.830 & 0.830 & 1 & 0.225 & 2.919 & 0.845 & - & - & {77.211} & {52 (8)} & {129} \\
6.0 & 25 & 0.935 & 0.935 & 1 & 0.219 & 3.351 & 0.946 & 4 & 18 & {54.590} & {48 (12)} & {103} \\
\end{tabular}
\end{table*}

\begin{table*}
\caption{Same as Table~\ref{t:4}, but for mixtures V and VI
  (Table~\ref{t:1})   with $Z=0.008$.}
\label{ta:4} 
\begin{tabular}{ccccccccccrrr}
\hline
\noalign{\smallskip}
$M_{\mathrm{ZAMS}}$ & No & $M_\mathrm{c}$(1) & $M_\mathrm{c}$(3du)
& No & C/O$_{\mathrm{f}}$ & $M$(f) & $M_\mathrm{c}$(f)
& No & No & \multicolumn{3}{c}{TP-AGB lifetime (10$^{3}$ yrs)} \\
\cline{11-13}
& TPs & & & TP$_{\mathrm{3du}}$(i) & & & & TP$_{\mathrm{HBB}}$(i) &
TP$_{\mathrm{HBB}}$(f) & $t_{\mathrm{mod}}$ & $\approx t_{\mathrm{e}}$
(TPs) & $t_{\mathrm{TP}}$  \\
\noalign{\smallskip}
\hline
\noalign{\smallskip}
\multicolumn{13}{c}{$Z = 0.008$ / solar (mixture V)} \\
\noalign{\smallskip}
\hline
\noalign{\smallskip}
1.0 & 5 & 0.513 & 0.513 & 1 & 1.170 & 0.619 & 0.531 & - & - & {698.898} & {85 (0)} & {784} \\
1.2 & 4 & 0.521 & 0.521 & 1 & 1.309 & 0.663 & 0.531 & - & - & {482.655} & {41 (0)} & {524} \\
1.5 & 5 & 0.521 & 0.524 & 2 & 2.204 & 0.538 & 0.537 & - & - & {794.177} & {-} & {794} \\
1.6 & 7 & 0.516 & 0.520 & 2 & 2.570 & 0.555 & 0.537 & - & - & {1029.957} & {4 (0)} & {1034} \\
1.8 & 9 & 0.497 & 0.499 & 2 & 2.004 & 1.164 & 0.526 & - & - & {1556.110} & {201 (1)} & {1757} \\
2.0 & 16 & 0.464 & 0.464 & 1 & 1.732 & 1.354 & 0.528 & - & - & {4081.799} & {260 (1)} & {4342} \\
2.6 & 10 & 0.569 & 0.569 & 1 & 2.860 & 0.758 & 0.576 & - & - & {984.120} & {6 (0)} & {990} \\
3.0 & 11 & 0.672 & 0.672 & 1 & 1.699 & 2.133 & 0.686 & - & - & {352.172} & {144 (3)} & {496} \\
4.0 & 19 & 0.801 & 0.801 & 1 & 1.358 & 3.348 & 0.816 & - & - & {178.647} & {253 (25)} & {432} \\
5.0 & 38 & 0.868 & 0.868 & 1 & 1.062 & 2.443 & 0.879 & 9 & 24 & {188.148} & {62 (12)} & {250} \\
6.0 & 46 & 0.991 & 0.991 & 1 & 0.281 & 3.498 & 0.993 & 1 & 46 & {97.485} & {79 (39)} & {176} \\
\noalign{\smallskip}
\hline\noalign{\smallskip}
\multicolumn{13}{c}{$Z = 0.008$ / $\alpha$-enhanced (mixture VI)} \\
\noalign{\smallskip}
\hline
\noalign{\smallskip}
1.0 & 5 & 0.508 & 0.515 & 3 & 2.354 & 0.550 & 0.523 & - & - & {667.889} & {24 (0)} & {692} \\
1.2 & 5 & 0.517 & 0.520 & 2 & 2.434 & 0.554 & 0.531 & - & - & {641.089} & {6 (0)} & {647} \\
1.5 & 4 & 0.527 & 0.527 & 1 & 1.787 & 1.034 & 0.537 & - & - & {634.407} & {50 (0)} & {684} \\
1.6 & 5 & 0.522 & 0.522 & 1 & 1.999 & 1.173 & 0.536 & - & - & {830.354} & {63 (0)} & {893} \\
1.8 & 5 & 0.505 & 0.505 & 1 & 2.365 & 0.524 & 0.522 & - & - & {1446.966} & {-} & {1447} \\
2.0 & 12 & 0.464 & 0.472 & 2 & 1.916 & 1.513 & 0.498 & - & - & {3458.889} & {320 (1)} & {3779} \\
2.6 & 10 & 0.578 & 0.578 & 1 & 2.530 & 0.772 & 0.592 & - & - & {921.296} & {6 (0)} & {927} \\
3.0 & 12 & 0.684 & 0.684 & 1 & 1.526 & 2.074 & 0.698 & - & - & {367.321} & {87 (2)} & {454} \\
5.0 & 32 & 0.875 & 0.875 & 1 & 0.563 & 1.824 & 0.887 & 9 & 25 & {195.952} & {37 (6)} & {233} \\
6.0 & 44 & 0.996 & 0.996 & 2 & 0.508 & 3.089 & 0.988 & 1 & 42 & {108.084} & {66 (22)} & {174} \\
\end{tabular}
\end{table*}

\begin{table*}
\caption{Same as Table~\ref{t:4}, but for mixtures VII and VIII
(Table~\ref{t:1})   with $Z=0.004$.}
\label{ta:5} 
\begin{tabular}{ccccccccccrrr}
\hline
\noalign{\smallskip}
$M_{\mathrm{ZAMS}}$ & No & $M_\mathrm{c}$(1) & $M_\mathrm{c}$(3du)
& No & C/O$_{\mathrm{f}}$ & $M$(f) & $M_\mathrm{c}$(f)
& No & No & \multicolumn{3}{c}{TP-AGB lifetime (10$^{3}$ yrs)} \\
\cline{11-13}
& TPs & & & TP$_{\mathrm{3du}}$(i) & & & & TP$_{\mathrm{HBB}}$(i) &
TP$_{\mathrm{HBB}}$(f) & $t_{\mathrm{mod}}$ & $\approx t_{\mathrm{e}}$
(TPs) & $t_{\mathrm{TP}}$  \\
\noalign{\smallskip}
\hline
\noalign{\smallskip}
\multicolumn{13}{c}{$Z = 0.004$ / solar (mixture VII)} \\
\noalign{\smallskip}
\hline
\noalign{\smallskip}
1.0 & 4 & 0.517 & 0.522 & 2 & 4.279 & 0.537 & 0.531 & - & - & {590.409} & {3 (0)} & {593} \\
1.2 & 4 & 0.525 & 0.529 & 2 & 4.305 & 0.555 & 0.536 & - & - & {530.145} & {5 (0)} & {535} \\
1.5 & 5 & 0.527 & 0.531 & 2 & 3.581 & 0.731 & 0.537 & - & - & {691.108} & {19 (0)} & {710} \\
1.6 & 7 & 0.521 & 0.525 & 2 & 3.156 & 0.537 & 0.536 & - & - & {1056.562} & {-} & {1057} \\
1.8 & 5 & 0.499 & 0.505 & 2 & 3.318 & 0.598 & 0.493 & - & - & {1686.695} & {33 (0)} & {1720} \\
2.0 & 14 & 0.488 & 0.495 & 2 & 3.035 & 0.529 & 0.528 & - & - & {3609.374} & {-} & {3610} \\
2.6 &  9 & 0.639 & 0.639 & 1 & 2.868 & 0.914 & 0.653 & - & - & {463.827} & {8 (0)} & {472} \\
3.0 & 11 & 0.744 & 0.744 & 1 & 2.235 & 1.381 & 0.746 & - & - & {200.714} & {16 (0)} & {217} \\
4.0 & 26 & 0.809 & 0.809 & 1 & 2.442 & 1.655 & 0.818 & - & - & {259.943} & {2 (0)} & {263} \\
5.0 & 35 & 0.894 & 0.894 & 1 & 1.036 & 2.787 & 0.898 & 6 & 31 & {164.360} & {75 (15)} & {239} \\
6.0 & 62 & 1.060 & 1.060 & 1 & 0.707 & 3.461 & 1.047 & 1 & 62 & {88.046} & {2 (30)} & {90} \\
\noalign{\smallskip}
\hline\noalign{\smallskip}
\multicolumn{13}{c}{$Z = 0.004$ / $\alpha$-enhanced (mixture VIII)} \\
\noalign{\smallskip}
\hline
\noalign{\smallskip}
1.0 & 5 & 0.515 & 0.515 & 1 & 3.175 & 0.554 & 0.532 & - & - & {682.685} & {19 (0)} & {702} \\
1.2 & 5 & 0.523 & 0.523 & 1 & 4.177 & 0.538 & 0.533 & - & - & {674.829} & {0.6 (0)} &{675} \\
1.5 & 5 & 0.531 & 0.531 & 1 & 3.365 & 0.550 & 0.543 & - & - & {783.520} & {-} & {784} \\
1.6 & 3 & 0.529 & 0.529 & 1 & 2.752 & 0.521 & 0.519 & - & - & {894.331} & {-} & {894} \\
1.8 & 8 & 0.506 & 0.506 & 1 & 3.123 & 0.529 & 0.526 & - & - & {1529.455} & {-} & {1529} \\
2.0 & 10 & 0.498 & 0.498 & 2 & 3.449 & 0.519 & 0.512 & - & - & {2199.189} & {0.2 (0)} & {2200} \\
2.6 & 10 & 0.647 & 0.647 & 1 & 2.522 & 1.632 & 0.665 & - & - & {443.615} & {61 (1)} & {505} \\
5.0 & 38 & 0.908 & 0.908 & 1 & 0.772 & 2.642 & 0.912 & 4 & 32 & {159.533} & {69 (13)} & {229} \\
6.0 & 70 & 1.050 & 1.050 & 1 & 0.372 & 3.366 & 1.048 & 1 & 70 & {100.072} & {73 (36)} & {173} \\
\end{tabular}
\end{table*}

\begin{table*}
\caption{Same as Table~\ref{t:4}, but for mixtures IX and X
  (Table~\ref{t:1})   with $Z=0.0005$.}
\label{ta:6} 
\begin{tabular}{ccccccccccrrr}
\hline
\noalign{\smallskip}
$M_{\mathrm{ZAMS}}$ & No & $M_\mathrm{c}$(1) & $M_\mathrm{c}$(3du)
& No & C/O$_{\mathrm{f}}$ & $M$(f) & $M_\mathrm{c}$(f)
& No & No & \multicolumn{3}{c}{TP-AGB lifetime (10$^{3}$ yrs)} \\
\cline{11-13}
& TPs & & & TP$_{\mathrm{3du}}$(i) & & & & TP$_{\mathrm{HBB}}$(i) &
TP$_{\mathrm{HBB}}$(f) & $t_{\mathrm{mod}}$ & $\approx t_{\mathrm{e}}$
(TPs) & $t_{\mathrm{TP}}$  \\
\noalign{\smallskip}
\hline
\noalign{\smallskip}
\multicolumn{13}{c}{$Z = 0.0005$ / solar (mixture IX)} \\
\noalign{\smallskip}
\hline
\noalign{\smallskip}
1.0 & 4 & 0.523 & 0.527 & 2 & 5.184 & 0.539 & 0.537 & - & - & {600.331} & {-} & {600} \\
1.2 & 4 & 0.537 & 0.548 & 3 & 4.125 & 0.551 & 0.550 & - & - & {518.908} & {-} & {519} \\
1.5 & 4 & 0.570 & 0.582 & 3 & 4.219 & 0.579 & 0.578 & - & - & {553.484} & {-} & {553} \\
1.6 & 5 & 0.561 & 0.566 & 2 & 4.462 & 0.579 & 0.578 & - & - & {619.811} & {-} & {620} \\
1.8 & 5 & 0.551 & 0.551 & 1 & 4.205 & 0.565 & 0.564 & - & - & {733.834} & {-} & {734} \\
2.0 & 6 & 0.584 & 0.588 & 2 & 4.180 & 0.600 & 0.599 & - & - & {603.197} & {-} & {603} \\
2.6 & 9 & 0.733 & 0.733 & 1 & 3.560 & 1.019 & 0.734 & - & - & {211.692} & {3 (0)} & {215} \\
3.0 & 10 & 0.780 & 0.780 & 1 & 3.776 & 1.277 & 0.772 & - & - & {162.996} & {25 (1)} & {188} \\
4.0 & 20 & 0.848 & 0.848 & 2 & 1.861 & 2.397 & 0.829 & - & - & {178.286} & {31 (3)} & {209} \\
\noalign{\smallskip}
\hline
\multicolumn{13}{c}{$Z = 0.0005$ / $\alpha$-enhanced (mixture X)} \\
\noalign{\smallskip}
\hline
\noalign{\smallskip}
1.0 & 3 & 0.528 & 0.528 & 1 & 6.183 & 0.535 & 0.533 & - & - & {446.329} & {-} & {446} \\
1.2 & 3 & 0.538 & 0.538 & 1 & 6.590 & 0.547 & 0.545 & - & - & {483.113} & {-} & {483} \\
1.5 & 4 & 0.560 & 0.560 & 1 & 5.602 & 0.568 & 0.567 & - & - & {510.437} & {-} & {510} \\
1.6 & 3 & 0.564 & 0.564 & 1 & 2.568 & 0.559 & 0.558 & - & - & {456.902} & {-} & {457} \\
1.8 & 6 & 0.555 & 0.555 & 1 & 4.588 & 0.561 & 0.561 & - & - & {752.815} & {-} & {753} \\
2.0 & 6 & 0.588 & 0.588 & 1 & 4.716 & 0.599 & 0.598 & - & - & {577.109} & {-} & {577} \\
\end{tabular}
\end{table*}

\clearpage

\section{Post-AGB evolution}

\begin{table*}
\caption{Post-AGB evolution for all models with solar-scaled
  metal distributions computed through this phase.$^a$ }
\label{ta:7} 
\begin{tabular}{ccccccccc}
\hline \noalign{\smallskip}
$M_{\mathrm{ZAMS}}$ & $T_{\mathrm{eff,i}}$ & $M_{\mathrm{T4}}$ &
$t_{\mathrm{tr}}$ & $t_\mathrm{1}$ & $t_\mathrm{cr}$ &?-burner & TP & End \\
\noalign{\smallskip} 
\hline\noalign{\smallskip}
\multicolumn{9}{c}{Z = 0.0005 / solar} \\
\noalign{\smallskip} \hline \noalign{\smallskip}
1.0 & 3.68 & 0.537 & 4.274 & 5.365 & 14.464 & H & no & WD \\
1.2 & 3.69 & 0.550 & 2.403 & 2.868 & 8.845  & H & no & WD \\
1.5 & 3.72 & 0.579 & 0.905 & 1.008 & 3.441  & H & no & WD \\
1.6 & 3.72 & 0.578 & 1.076 & 1.213 & 4.129  & H & no & WD \\
1.8 & 3.71 & 0.565 & 0.545 & 0.860 & 3.917  & H & no & WD \\
2.0 & 3.73 & 0.600 & 0.335 & 0.478 & 2.412  & H & no & WD \\
\noalign{\smallskip} \hline \noalign{\smallskip}
\multicolumn{9}{c}{Z = 0.004 / solar} \\
\noalign{\smallskip} \hline \noalign{\smallskip}
1.0 & 3.71* & 0.528 & 0.030 & 1.116* & 16.348 & H    & no   & WD \\
1.2 & 3.71* & 0.531 & 0.032 & 1.386* & 16.816 & H    & no   & W  \\
1.5 & 3.71* & 0.536 & 0.276 & 0.681* & 8.149  & H/He & VLTP & 2$^\mathrm{nd}$ AGB \\
1.6 & 3.69  & 0.536 & 1.605 & 1.909  & 6.330  & H    & no   & WD  \\
1.8 & 3.71* & 0.492 & 0.020 & 1.029* & 2.810  & H/He & LTP  & 2$^\mathrm{nd}$ AGB \\
2.0 & 3.68  & 0.528 & 1.325 & 2.770 & 10.709  & H    & no   & WD   \\
\noalign{\smallskip} \hline \noalign{\smallskip}
\multicolumn{9}{c}{Z = 0.008 / solar} \\
\noalign{\smallskip} \hline \noalign{\smallskip}
1.0 & 3.70* & 0.531 & 0.577 & 1.730* & 16.935 & H/He & VLTP & 2$^\mathrm{nd}$ AGB \\
1.2 & 3.71* & 0.529 & 0.027 & 0.030* & 14.369 & H & no & WD \\
1.5 & 3.69  & 0.537 & 1.840 & 2.178  & 6.995  & H/He & VLTP & during VLTP \\
1.6 & 3.72* & 0.537 & 0.192 & 0.443* & 5.595  & H/He & VLTP & 2$^\mathrm{nd}$ AGB \\
1.8 & 3.72* & 0.526 & 0.021 & 0.025* & 12.811 & H & no & WD \\
2.0 & 3.73* & 0.528 & 0.013 & 0.014* & 5.173  & H & no & WD \\
\noalign{\smallskip} \hline \noalign{\smallskip}
\multicolumn{9}{c}{Z = 0.02 / solar} \\
\noalign{\smallskip} \hline \noalign{\smallskip}
1.0 & 3.71* & 0.506 & 0.037 & 0.041* & 26.417 & H & no & WD \\
1.2 & 3.70* & 0.523 & 0.033 & 0.038* & 16.099 & H & no & WD \\
1.5 & 3.70* & 0.538 & 0.445 & 0.832* & 7.388  & H & no & WD \\
1.6 & 3.72* & 0.541 & 0.020 & 0.021* & 6.248  & H/He & VLTP & 2$^\mathrm{nd}$ AGB \\
1.8 & 3.73* & 0.524 & 0.667 & 29.316*& 47.833 & H & no & WD \\
2.0 & 3.70  & 0.543 & 0.528 & 0.868  & 3.614  & H & no & WD \\
\noalign{\smallskip} \hline \noalign{\smallskip}
\multicolumn{9}{c}{Z = 0.04 / solar} \\
\noalign{\smallskip} \hline \noalign{\smallskip}
1.0 & 3.57  & 49.717 & 0.510 & 57.677 & 86.516 & He & no & WD \\
1.2 & 3.60  & 4.042  & 0.522 & 31.515 & 48.850 & He & no & WD \\
1.5 & 3.64  & 7.858  & 0.531 & 10.215 & 18.553 & H & no & WD \\
1.6 & 3.71* & 0.021  & 0.536 & 0.029* & 6.926  & H & no & WD \\
1.8 & 3.70* & 0.783  & 0.536 & 1.805* & 6.875  & H & no & WD \\
2.0 & 3.70  & 0.628  & 0.552 & 1.131  & 4.523  & H & no & WD \\
\noalign{\smallskip} \hline 
\noalign{\medskip
$^a$ see text for explanation of column contents; all times in  $10^3$~years.}
\end{tabular}
\end{table*}

\begin{table*}
\caption{As Table~\ref{ta:7}, but for the $\alpha$-enhanced composition
  tracks.}
\label{ta:8} 
\begin{tabular}{ccccccccc}
\hline \noalign{\smallskip}
$M_{\mathrm{ZAMS}}$ & $T_{\mathrm{eff,i}}$ & $M_{\mathrm{T4}}$ &
$t_{\mathrm{tr}}$ & $t_\mathrm{1}$ & $t_\mathrm{cr}$ &?-burner & TP & End \\
\noalign{\smallskip} \hline \noalign{\smallskip}
\multicolumn{9}{c}{Z = 0.0005 / $\alpha$-enhanced} \\
\noalign{\smallskip} \hline \noalign{\smallskip}
1.0 & 3.67 & 0.533 & 4.880 & 6.219 & 16.959& H & no & WD \\
1.2 & 3.69 & 0.545 & 2.742 & 3.431 & 10.276& H & no & WD \\
1.5 & 3.71 & 0.567 & 1.341 & 1.585 & 5.490 & H & no & WD \\
1.6 & 3.71 & 0.558 & 1.248 & 1.438 & 4.958 & H/He & VLTP & 2$^\mathrm{nd}$ AGB \\
1.8 & 3.71 & 0.561 & 0.575 & 1.008 & 4.649 & H & no & WD \\
2.0 & 3.73 & 0.598 & 0.332 & 0.510 & 3.621 & H & no & WD \\
\noalign{\smallskip} \hline \noalign{\smallskip}
\multicolumn{9}{c}{Z = 0.004 / $\alpha$-enhanced} \\
\noalign{\smallskip} \hline \noalign{\smallskip}
1.0 & 3.71* & 0.530 & 3.169 & 4.600* & 17.434& H & no & WD \\
1.2 & 3.72* & 0.538 & 1.985 & 2.690* & 11.122& H/He & VLTP & 2$^\mathrm{nd}$ AGB \\
1.5 & 3.70  & 0.543 & 1.725 & 2.062  & 6.962 & H & no & WD \\
1.6 & 3.69  & 0.519 & 1.636 & 1.969  & 3.728 & H/He & VLTP & 2$^\mathrm{nd}$ AGB  \\
1.8 & 3.67  & 0.527 & 1.756 & 4.160  & 13.545& H & no & WD \\
2.0 & 3.73  & 0.515 & 0.510 & 1.014  & 0.016 & H/He & LTP & 2$^\mathrm{nd}$ AGB \\
\noalign{\smallskip} \hline \noalign{\smallskip}
\multicolumn{9}{c}{Z = 0.008 / $\alpha$-enhanced} \\
\noalign{\smallskip} \hline \noalign{\smallskip}
1.0 & 3.71* & 0.524 & 2.572 & 3.599* & 1.141 & H/He & LTP & 2$^\mathrm{nd}$ AGB \\
1.2 & 3.72* & 0.531 & 0.027 & 0.732* & 13.983& H/He & VLTP & 2$^\mathrm{nd}$ AGB \\
1.5 & 3.73* & 0.537 & 0.016 & 0.016* & 0.859 & H/He & VLTP & 2$^\mathrm{nd}$ AGB \\
1.6 & 3.74* & 0.536 & 0.013 & 0.014* & 6.688 & H/He & VLTP & 2$^\mathrm{nd}$ AGB \\
1.8 & 3.68  & 0.523 & 1.039 & 2.062  & 8.281 & H & no & WD \\
2.0 & 3.73* & 0.498 & 0.012 & 0.013* & 6.163 & H & no & WD \\
\noalign{\smallskip} \hline \noalign{\smallskip}
\multicolumn{9}{c}{Z = 0.02 / $\alpha$-enhanced} \\
\noalign{\smallskip} \hline \noalign{\smallskip}
1.0 & 3.71* & 0.508 & 0.046 & 2.798* & 16.572& H/He & LTP & 2$^\mathrm{nd}$ AGB \\
1.2 & 3.72* & 0.519 & 0.026 & 0.027* & 14.592& H & no & WD \\
1.5 & 3.71* & 0.537 & 1.498 & 1.833* & 7.137 & H & no & WD \\
1.6 & 3.70  & 0.548 & 1.368 & 1.594  & 2.903 & H/He & LTP & 2$^\mathrm{nd}$ AGB \\
1.8 & 3.73* & 0.540 & 0.495 & 0.973* & 0.844 & H/He & LTP & 2$^\mathrm{nd}$ AGB \\
2.0 & 3.71* & 0.537 & 0.113 & 0.403* & 3.827 & H & no & WD \\
\noalign{\smallskip} \hline \noalign{\smallskip}
\multicolumn{9}{c}{Z = 0.04 / $\alpha$-enhanced} \\
\noalign{\smallskip} \hline \noalign{\smallskip}
1.0 & 3.58 & 0.511 & 51.989 & 58.338 & 35.345 & He/He & LTP & 2$^\mathrm{nd}$ AGB \\
1.2 & 3.63 & 0.521 & 10.657 & 14.043 & 25.677 & H & no & WD \\
1.5 & 3.60 & 0.536 & 23.111 & 16.080 & 0.021  & He & no & WD \\
1.6 & 3.72*& 0.535 & 24.661 & 0.021* & 7.894  & H/He & VLTP & 2$^\mathrm{nd}$ AGB \\
1.8 & 3.66 & 0.537 & 2.417  & 5.980  & 16.977 & H & no & WD \\
2.0 & 3.73*& 0.552 & 0.539  & 0.885* & 0.387  & H/He & LTP & 2$^\mathrm{nd}$ AGB \\
\noalign{\smallskip} \hline \noalign{\smallskip}
\end{tabular}
\end{table*}

\end{appendix}

\end{document}